\RequirePackage{fix-cm}
\documentclass[natbib,smallextended]{svjour3}       % onecolumn (second format)
\smartqed  % flush right qed marks, e.g. at end of proof
\usepackage{graphicx}
\usepackage{amssymb,amsmath}
\usepackage{hyperref}
\hypersetup{
    colorlinks,%
    citecolor=blue,%
    linkcolor=blue,%
    urlcolor=blue
}

 \journalname{my journal}
\def\bE{{\mathbf E}}
\def\bB{{\mathbf B}}
\def\bV{{\mathbf V}}
\def\bv{{\mathbf v}}
\def\bJ{{\mathbf J}}
\def\bOm{\mathbf{\Omega}}
\def\bmu{\boldsymbol{\mu}}
\def\br{{\mathbf r}}
\def\rs{r_\star}
\def\RLC{R_{\rm LC}}
\def\rhoGJ{\rho_{\rm GJ}}

\def\Eq{Equation}
\def\Eqs{Equations}

\newcommand\beq{\begin{equation}}
\newcommand\eeq{\end{equation}}

\begin{document}

\title{Electrodynamics of pulsar magnetospheres}

%\subtitle{}

\author{Beno\^it Cerutti  \and
        Andrei M. Beloborodov %etc.
}


\institute{B. Cerutti \at
              Univ. Grenoble Alpes, CNRS, IPAG, F-38000 Grenoble, France.\\
              Tel: +33476514874\\
              Fax: +33476448821\\
              \email{benoit.cerutti@univ-grenoble-alpes.fr}           %  \\
           \and
           A. M. Beloborodov \at
              Physics Department and Columbia Astrophysics Laboratory, Columbia University, 538 West 120th Street, New York, NY 10027, USA.
              \email{amb@phys.columbia.edu}           %  \\
}

\date{Received: date / Accepted: date}

\maketitle

\begin{abstract}

We review electrodynamics of rotating magnetized neutron stars, from the early vacuum model to recent numerical experiments with plasma-filled magnetospheres. Significant progress became possible due to the development of global particle-in-cell simulations which capture particle acceleration, emission of high-energy photons, and electron-positron pair creation. The numerical experiments show from first principles how and where electric gaps form, and promise to explain the observed pulsar activity from radio waves to gamma-rays.

\end{abstract}

% B_{\rm r} -> B_r

\section{Introduction}\label{sect_intro}

The canonical radio pulsar is an isolated, rotating, magnetized neutron star. Its radius is $r_{\star}=10$--13~km and its mass is $M_{\star}\approx 1.4M_\odot$. The rotation periods of pulsars vary from milliseconds to seconds and their surface magnetic fields --- from $10^9$~G to $10^{14}$~G. Pulsars have been a great puzzle since their discovery in 1967 \citep{1968Natur.217..709H}. It is established that the canonical pulsars are powered by the kinetic energy of their rotation. Their powerful electromagnetic winds extract the rotational energy of the star and create bright nebulae, e.g. the famous Crab nebula. More surprisingly, pulsars themselves are bright sources of broad-band radiation, from radio waves to gamma-rays, whose mechanisms have been debated for almost five decades.

The precise measurements of the pulsar rotation period $P$ and its time derivative $\dot{P}$ provide accurate knowledge of the energy reservoir and the total released power. A significant fraction of the spindown power (typically 1-10$\%$) is converted to high-energy gamma-rays, as clearly demonstrated by the {\em Fermi}-LAT observations in the 0.1-10~GeV band \citep{2010ApJS..187..460A, 2013ApJS..208...17A}. Thus pulsars must be efficient particle accelerators. Understanding their mechanism is a difficult task, because it involves a complex interplay between electrodynamics, high-energy radiative processes, and electron-positron pair creation.

The basic picture of pulsar magnetosphere needs to be established before addressing the many puzzles posed by pulsar observations --- the extremely high number of $e^\pm$ ejected by the Crab pulsar, the drifting radio sub-pulses, the giant radio pulses, intermittency, gamma-ray pulse profiles, the spectrum shape etc. In this introductory review, we summarize the attempts to understand the basic picture of the pulsar magnetosphere over the past five decades. The topic has been discussed in many reviews, e.g. by \citet{1979SSRv...24..437A}; \citet{2004AdSpR..33..542M} and recently by \citet{2009ASSL..357..373A, 2012SSRv..173..341A, 2011ASSP...21..139S, 2014RPPh...77f6901B, 2015SSRv..191..207B, 2015CRPhy..16..641G, 2016arXiv160804895P}. We focus on the pulsar itself, i.e. the region inside and around the light cylinder where the pulsar wind forms; an excellent theoretical review of pulsar-wind nebulae is found in \citet{2009ASSL..357..421K}.

\section{Vacuum dipole magnetosphere}\label{sect_vacuum}

In the early model of pulsar magnetosphere, the star is surrounded by vacuum \citep{1967Natur.216..567P, 1968Natur.219..145P, 1969ApJ...157.1395O}. The initial argument behind this assumption was that the strong gravity of the neutron star keeps any free particles at the stellar surface and does not allow them to fill the magnetosphere \citep{1964Natur.203..914H}. This model neglects the lifting of particles by strong induced electric fields and pair creation in the magnetosphere, which will be discussed in the next sections.

The pulsar is described by its magnetic moment $\boldsymbol{\mu}$ and angular velocity $\boldsymbol{\Omega}$, and the light cylinder radius is defined by $R_{\rm LC}=c/\Omega$. Let us first consider the ``aligned rotator'' $\boldsymbol{\mu} \parallel \boldsymbol{\Omega}$ with the magnetic axis passing through the center of the star. Clearly, this ``pulsar'' would not pulsate, as the configuration is symmetric about the rotation axis. However, understanding this simplest configuration plays an important role in the development of pulsar theory.

The neutron star is an excellent conductor, and therefore the electric field measured in the corotating frame inside the star nearly vanishes, $\bE^\prime_{\rm int}=\mathbf{0}$. This condition determines the internal electric field in the fixed lab frame $E_{\rm int}$,
\begin{equation}
\bE^\prime_{\rm int}=\mathbf{E_{\rm int}}+\frac{\mathbf{V}\times\mathbf{B_{\rm int}}}{c}=\mathbf{0},
\qquad \bV=\mathbf{\Omega}\times\mathbf{r}.
\label{E_corotation}
\end{equation}
It implies that the star is polarized by rotation, very much like a Faraday disk (the unipolar induction effect). Electrons re-distribute themselves inside the star to create the electric field $\bE_{\rm int}$ that balances the Lorentz force $\bV\times\bB_{\rm int}/c$, where $\bV= c\mathbf{E_{\rm int}} \times \mathbf{B_{\rm int}}/B^2_{\rm int}=\bOm\times\br$ is the drift speed of the electrons. This polarization creates an excess of negative charges at the poles and positive charges at the equator.

\begin{figure}[]
\centering
\includegraphics[width=10cm]{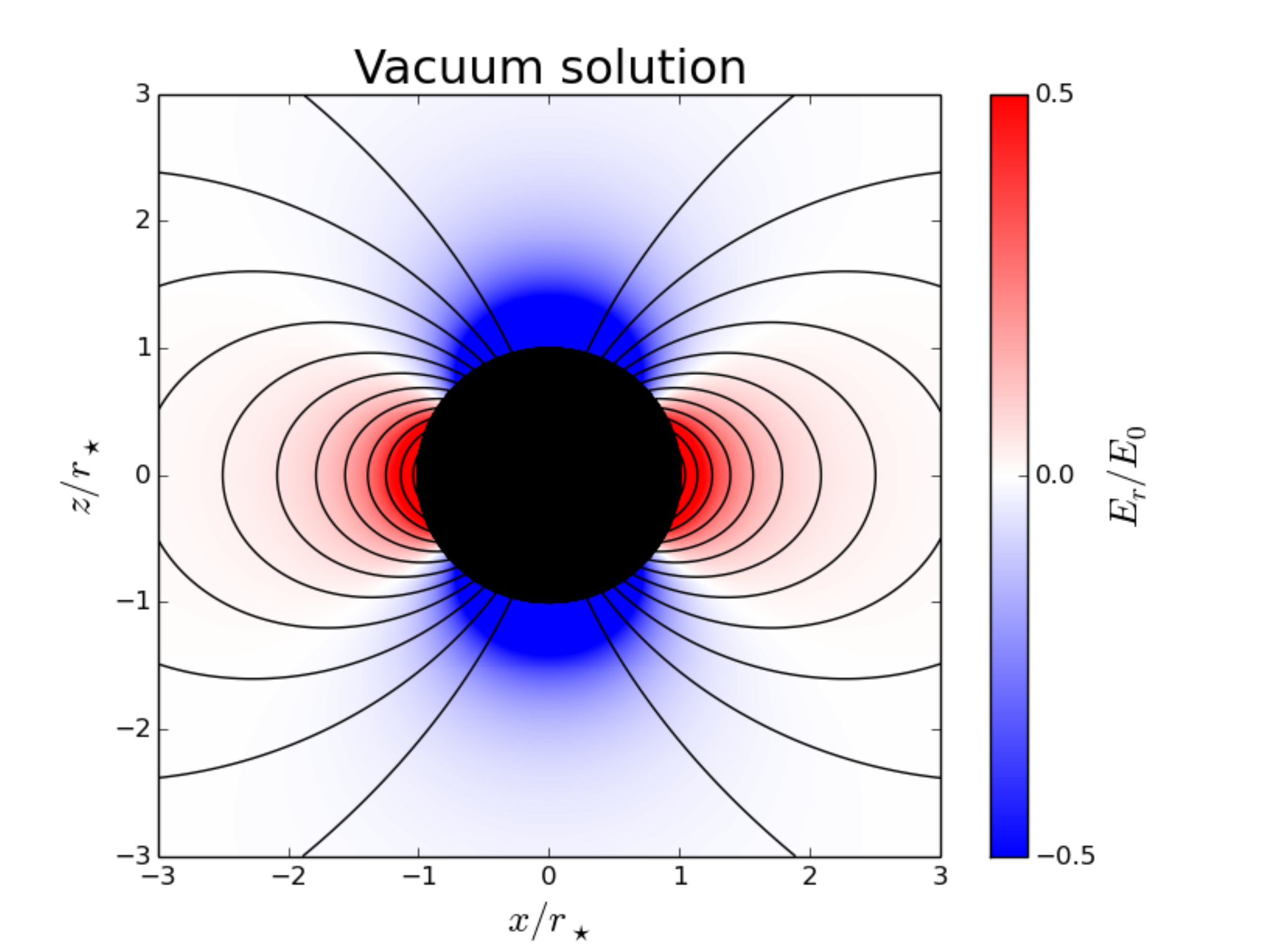}
\includegraphics[width=10cm]{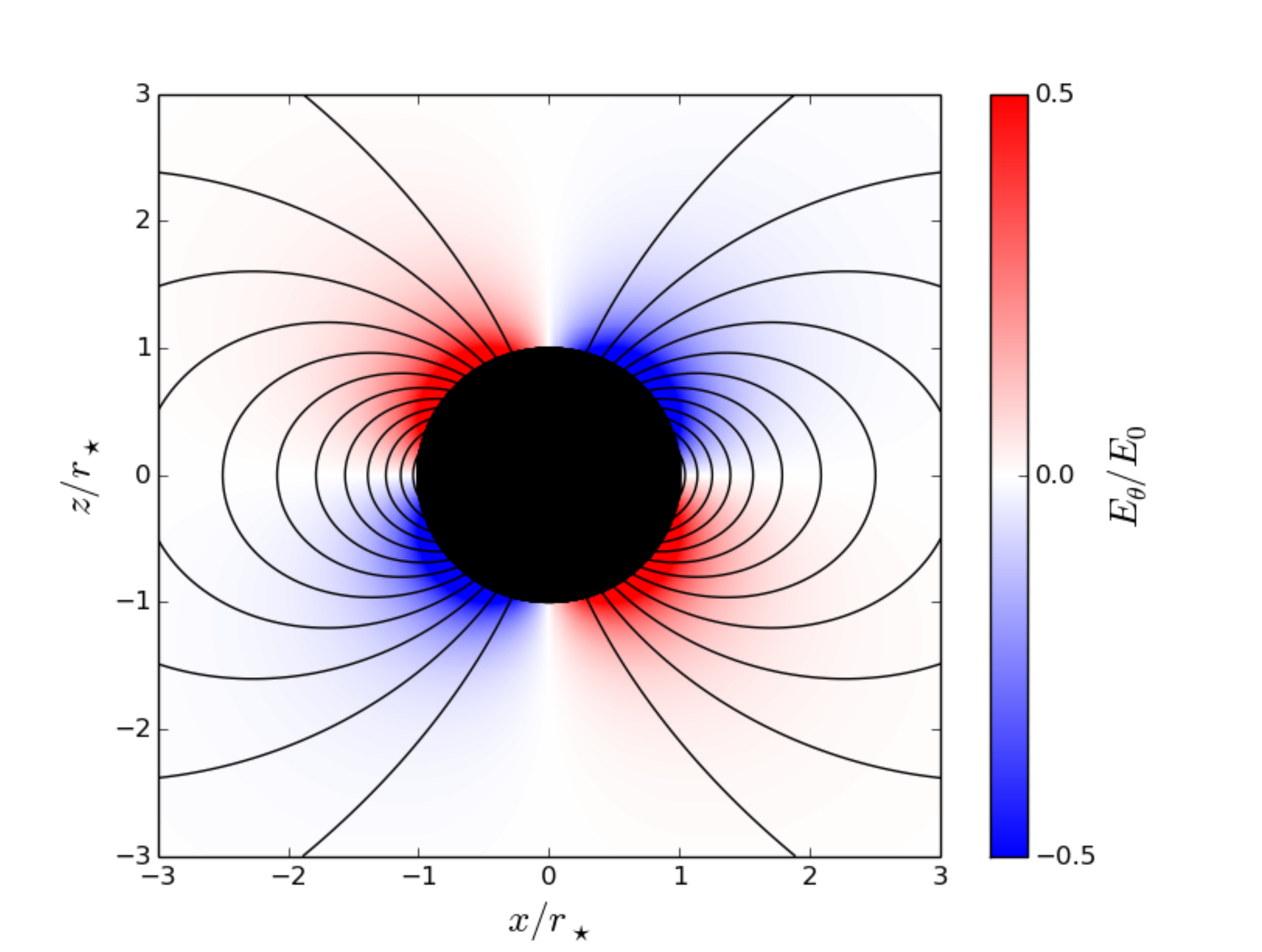}
\caption{Vacuum electromagnetic field of an aligned dipole rotator, $\bOm\parallel\bmu$ (both along the $z$-axis in the figure). Black curves show the magnetic field lines. Color shows the electric field outside the star (top: $E_{\rm r}$, bottom: $E_{\theta}$) induced by the rotation of the perfectly conducting neutron star with no net charge ($Q=0$). The electric field is shown in units of $E_0=\Omega \mu/\rs^2c=\Omega\rs B_\star/c$.}
\label{fig_vacuum}
\end{figure}

The internal electric field, in particular its $\theta$-component $E_\theta^{\rm int}=-V_\phi B_r^{\rm int}/c$ (we use spherical coordinates $r,\theta,\phi$) determines the electrostatic potential drop with latitude on the stellar surface $\Phi(\rs,\theta)-\Phi(\rs,0)=-\int E_\theta^{\rm int}\, \rs d\theta$. Using the surface dipole magnetic field $B_r=2\mu \cos\theta/\rs^3$ and $V_\phi = \Omega \rs \sin\theta$ one finds 
\begin{equation}
\label{eq:bc}
\Phi(\rs,\theta)=\Phi_p+\frac{\Omega\mu}{c\rs}\sin^2\theta,
\qquad \Phi_p
% =
\equiv
\Phi(\rs,0).
\end{equation}
It is straightforward to solve Laplace equation $\nabla^2\Phi=0$ in the vacuum domain outside the star with the boundary condition given in \Eq~(\ref{eq:bc}) (using that $\sin^2\theta-2/3=-(2/3)P_2(\theta)$ is an eigen function of the $\theta$-part of the Laplacian). The solution is given by \citep{1947PhRv...72..632D, 1955AnAp...18....1D, 1965JGR....70.4951H}
\beq
\Phi(r,\theta)=\left(\Phi_p+\frac{2\Omega\mu}{3c\rs}\right)\frac{\rs}{r}
+\frac{\Omega \mu}{c\rs}\left(\frac{\rs}{r}\right)^3\left(\sin^2\theta-\frac{2}{3}\right),
\qquad r\geq\rs.
\eeq
$\Phi_p$ determines the monopole contribution to $\Phi$. It is related to the net charge of the star $Q=-\rs\Phi_p-2\Omega\mu/3c$, as seen from the Gauss theorem applied to a sphere of radius $r\gg\rs$. If the neutron star is formed with zero charge and no particles are lifted from its surface then $Q=0$ and $\Phi_p=-(2\Omega\mu/3c\rs)$.

The electric field outside the star is found from $\bE=-\nabla\Phi$. Note that while $\Phi$ is continuous at the surface $r=\rs$, its derivative is not. Thus, a continuous dipolar magnetic field implies a discontinuous $E_r$ i.e. there is a surface charge $4\pi\Sigma=E_r-E_r^{\rm int}$. The star also possesses an interior charge, which is found from the Gauss theorem applied to a sphere just inside $\rs$ using $E_r^{\rm int}(\rs,\theta) =(V_\phi/c)B_\theta^{\rm int} =\Omega\mu\sin^2\theta/c\rs^2$, which gives $Q_{\rm int}=2\Omega\mu/3c$ \citep{1999PhR...318..227M}. The net charge $Q$ is the sum of the interior and surface charges.

The dipole magnetic field remains a valid vacuum solution for the aligned rotator, as if there were no rotation. In summary, the vacuum electromagnetic field of the aligned dipole rotator with $Q=0$ 
% (usually called the Deutsch field) 
is given by\footnote{Generalization of the Deutsch solution to an arbitrary multipolar order is discussed by \citet{2015A&A...573A..51B, 2015MNRAS.450..714P}.}
\begin{eqnarray}
 (B_r,B_\theta,B_\phi) &=& B_{\star}\left(\frac{r_{\star}}{r}\right)^3(2\cos\theta,\sin\theta,0)\\
 (E_r,E_\theta,E_\phi) &=& \frac{\Omega \rs}{c}B_{\star}\left(\frac{r_{\star}}{r}\right)^4 \left(1-3\cos^2\theta,-\sin 2\theta,0\right)
\end{eqnarray}
where $B_{\star}=\mu/\rs^3$ is the magnetic field strength at the stellar equator. The vacuum electric field has a monopole component if $Q\neq 0$; otherwise it is a pure quadrupole (see Figure~\ref{fig_vacuum}).

% The Deutsch solution determines the Poynting flux $c\,\bE\times\bB/4\pi$. 
The net Poynting flux from the aligned rotator (integrated over a sphere of radius $r>\rs$) vanishes, $L_{\rm vac}=0$. Therefore, the rotating star does not lose energy and hence does not spin down. The vacuum solution can be generalized to the case of an inclined rotator, where the magnetic moment is directed at an angle $\chi$ relative to the rotation axis. In this case, the net Poynting flux becomes \citep{1968Natur.219..145P, 1969ApJ...157.1395O}
\begin{equation}
   L_{\rm vac}=\frac{2}{3}\frac{\mu^2\Omega^4}{c^3}\sin^2\chi.
\label{spindown_vac}
\end{equation}

%##############################################################

\section{The electrosphere}\label{sect_elec}

Particles are strongly magnetized near the neutron star, so that their velocities perpendicular to the magnetic field lines are limited to the drift velocity $\bV_{D}=\bE\times\bB/B^2$. However, particles can freely slide along the magnetic field, in particular if there is a parallel electric field component $\mathbf{E_{\parallel}}=\mathbf{E}\cdot\mathbf{B}/|\mathbf{B}|\neq 0$. From the vacuum solution, one immediately realizes that there is a strong unscreened $E_\parallel$, in contrast to the interior of the star where $\mathbf{E}\cdot\mathbf{B}=0$. \citet{1969ApJ...157..869G} pointed out that the electric force greatly exceeds gravity of the star and will lift electrons in the polar region and ions in the equatorial region. The work function $W$ at the condensed surface of a neutron star crust is negligible for electrons and may be significant for ions \citep{1974IAUS...53..117R}. However, detailed calculations of $W$ \citep{1986MNRAS.218..477J, 2007MNRAS.382.1833M} show that the typical temperatures of pulsars provide sufficient surface emission of ions; these free ions are picked up by the electric field and lifted to the magnetosphere.

Thus the entire surface charge of the aligned rotator must be lifted and distributed in the magnetosphere. There exists an axisymmetric electrostatic equilibrium for the lifted charge \citep{1979ApJ...227..266J, 1985MNRAS.213P..43K, 1985A&A...144...72K}: the electrons lifted in the polar region form a dome above the star, and the ions lifted in the equatorial region form a torus. In both dome and torus $\bE\cdot\bB=0$. There is a vacuum gap between (and around) the dome and torus, i.e. the lifted charge does not fill the entire magnetosphere. The equilibrium solution for the lifted charge implies no outflow, no poloidal electric current, and no toroidal magnetic field $B_\phi$. This ``electrosphere'' produces no net Poynting flux and the pulsar is dead --- there is no spindown power.

In the corotating regions (stellar interior, dome, and part of the torus) the condition
$\bE=-\bV\times\bB/c$ determines charge density $\rho$ by 
$\nabla\cdot\bE=4\pi\rho$, which is often called ``corotation density'' or 
Goldreich-Julian density \citep{1965JGR....70.4951H, 1969ApJ...157..869G},
\begin{equation}
\rho_{\rm GJ}\approx-\frac{\mathbf{\Omega}\cdot\mathbf{B}}{2\pi c}=\frac{B_{\star}}{2\pi c}\left(\frac{r_{\star}}{r}\right)^3\left(3\cos^2\theta-1\right).
\label{density_GJ}
\end{equation}

%%%%%%%%%%%%%%%%%%%%%%%%%%%%%%%%%%%%%
\begin{figure}[t]
\centering
\includegraphics[width=10cm]{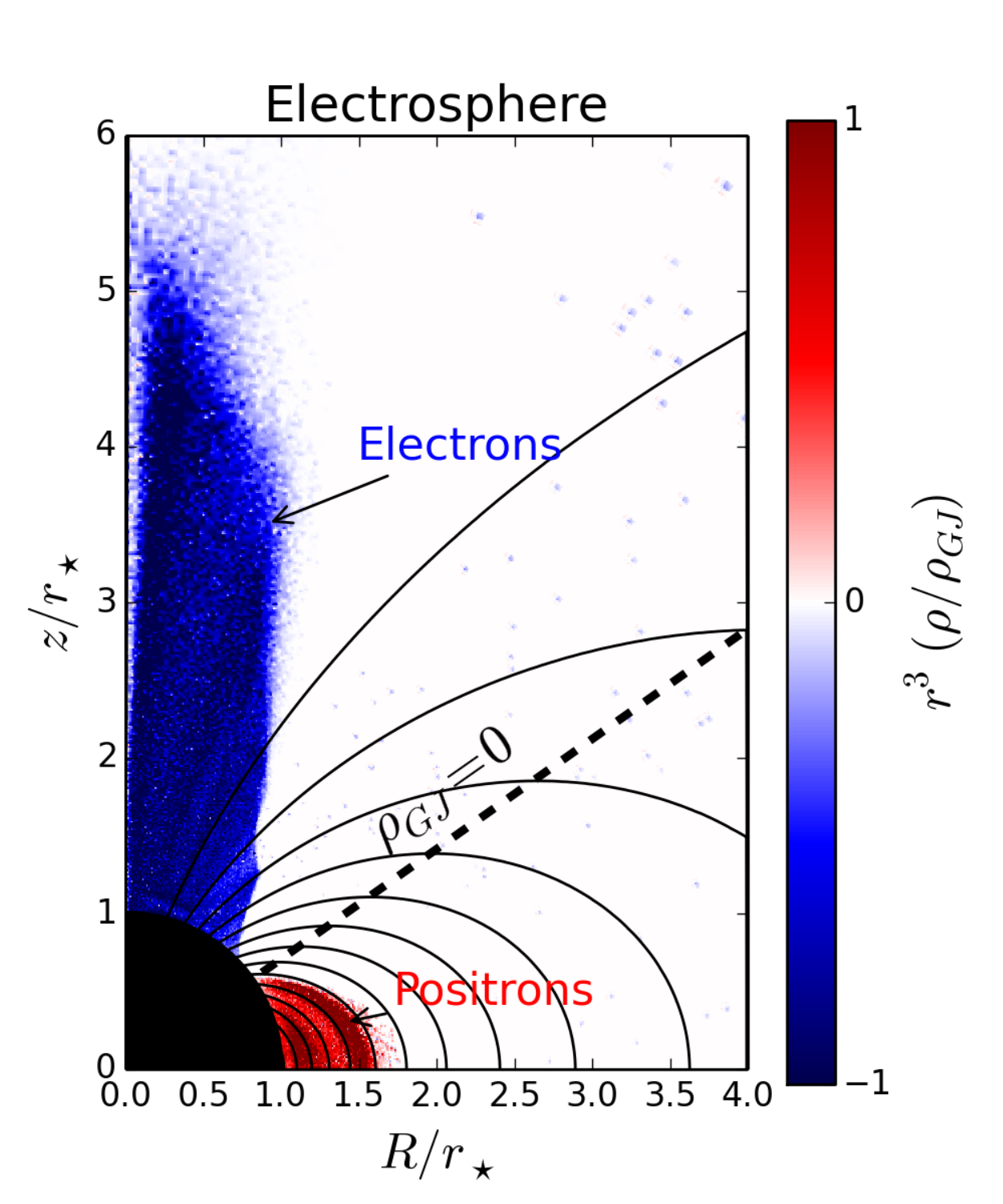}
\caption{Fully charge-separated solution (electrosphere) of the aligned rotator obtained with a 2D axisymmetric PIC simulation after one rotation period. Electrons (blue) form a dome above the magnetic pole while positive charges (here positrons, in red) form the equatorial torus. Both species are confined well within the light-cylinder radius, here set at $R_{\rm LC}=6 r_{\star}$. Charge densities are normalized by the surface Goldreich-Julian density at the pole. Solid curves show the magnetic field lines and the oblique dashed line shows the null line (where $\rho_{\rm GJ}=0$, i.e., $\theta\approx 55^{\rm o}$).}
\label{fig_electrosphere}
\end{figure}
%%%%%%%%%%%%%%%%%%%%%%%%%%%%%%%%%%%%%

The electrosphere configuration was obtained by several groups using iterative or PIC numerical simulations \citep{1985MNRAS.213P..43K, 1985A&A...144...72K, 1989Ap&SS.161..187S, 1993A&A...274..319N, 1994ApJ...431..718T, 2001MNRAS.322..209S, 2002A&A...384..414P, 2002ASPC..271...81S, 2009ApJ...690...13M, 2011MNRAS.418..612W, 2014ApJ...785L..33P, 2015MNRAS.448..606C}. In almost all cases, the simulations are initiated with the vacuum solution and zero work function at the stellar surface. Then, as expected, the electric field extracts charges from the star; they fill the magnetosphere and form the charge-separated dome+torus structure. The charge distribution obtained by an axisymmetric PIC simulation \citep{2015MNRAS.448..606C} is shown in Figure~\ref{fig_electrosphere}.

The equilibrium electrosphere model is however incomplete, because it turns out unstable to non-axisymmetric perturbations \citep{2002A&A...387..520P, 2002ASPC..271...81S}. The magnetosphere rotates with the drift velocity $\bE\times\bB/B^2$, and the electrosphere solution implies a strong velocity shear. This causes the diocotron instability of the torus, an analog of the Kelvin-Helmholtz instability in neutral fluids. This instability is captured only by 3D simulations as it grows from non-axisymmetric modes. It induces an expansion of the torus, and can even produce an outflow of charge through the light cylinder. However, this outflow is too weak to give a significant spindown power. Furthermore, \citet{2007A&A...469..843P} showed that the diocotron instability is suppressed by relativistic effects that become important near the light cylinder.

In summary, the electrosphere formed by particles lifted from the surface can hardly explain the spindown and the magnetospheric activity of pulsars. The model is, however, useful for old inactive neutron stars.

%#################################################################

\section{The force-free magnetosphere: The plasma-filled solution}\label{sect_filled}

Observations of pulsar wind nebulae indicate that the wind is heavily loaded with $e^\pm$ plasma, which must be created in the pulsar magnetosphere (e.g., \citealt{2009ASSL..357..421K}). Creation of $e^\pm$ pairs is also expected theoretically, due to strong electric fields that must develop in plasma-starved regions --- so-called ``gaps.'' For instance, the electrosphere solution has a gap between the dome and the torus, and a seed electron placed in the gap will be accelerated to enormous Lorentz factors. The electron is accelerated by $\mathbf{E}_{\parallel}$ along the curved field lines and emits high-energy photons (curvature radiation) which convert to $e^\pm$ pairs in the strong magnetic field \citep{1966RvMP...38..626E, 2006RPPh...69.2631H}. This process ignites an electromagnetic cascade of gamma rays and pairs until the density of the plasma is high enough to screen the accelerating electric field, so that $\mathbf{E}\cdot\mathbf{B}=0$ becomes nearly satisfied.

Although the existence of gaps is required to fill the magnetosphere with pair plasma, as a first approximation it makes sense to study ``force-free'' magnetospheres with $\bE\cdot\bB\approx 0$. This model assumes that small deviations from the screening condition are sufficient to fill the magnetosphere with plasma. The model also assumes that the inertial mass density of the plasma is much smaller than $B^2/8\pi c^2$. This limit is called ``force-free electrodynamics'' (FFE). It satisfies the equation,
\begin{equation}
\rho\mathbf{E}+\frac{\mathbf{J}\times\mathbf{B}}{c}=\mathbf{0},
\label{eq_ff}
\end{equation}
where $\rho$ and $\mathbf{J}$ are the charge and current densities. The FFE describes the behavior of the electromagnetic field, however does not provide any information about the plasma creation and dynamics, except that it sustains the electric current $\bJ$ and charge density $\rho=\nabla\cdot\bE/4\pi$ demanded by the electromagnetic field.

\subsection{Axisymmetric ``pulsar equation''}

Since the $\phi$-derivative vanishes for axisymmetric configurations, the condition $\nabla\cdot\bB=0$ gives $\nabla\cdot \bB_{\rm P}=0$. This condition implies that there is only one degree of freedom in the poloidal magnetic field $\bB_{\rm P}=(B_r,B_\theta)$. It is convenient to use the so-called ``poloidal flux function'' $\Psi(r,\theta)$ or $\Psi(R,z)$ in cylindrical coordinates $R,z$. It is defined so that $2\pi\Psi$ equals the magnetic flux through the circle of radius $R=r\sin\theta$ around the rotation axis at height $z=r\cos\theta$. The flux function $\Psi(R,z)$ is related to $\bB_{\rm P}$ by
\begin{equation}
\mathbf{B_{\rm P}}=\frac{\mathbf{\nabla}\Psi\times\mathbf{e_{\phi}}}{R}.
\end{equation}
The poloidal electric current through the same circle, $2\pi I$, is related to the toroidal magnetic field $B_\phi$ according to the Stokes' theorem,
\beq
   B_\phi=\frac{I}{R}.
\eeq 
The two scalar functions --- the flux function $\Psi(R,z)$ and the current function $I(R,z)$ --- completely describe an axisymmetric magnetic field.

In a steady state, it is possible to translate the force-free condition into the so-called ``pulsar equation'' for $\Psi$ and $I$ \citep{1973ApJ...182..951S, 1973ApJ...180L.133M}. This equation reads,
\begin{equation}
\left(1-\frac{R^2}{R^2_{\rm LC}}\right)\left(\frac{\partial^2\Psi}{\partial R^2}+\frac{\partial^2\Psi}{\partial z^2}\right)-\left(1+\frac{R^2}{R^2_{\rm LC}}\right)\frac{1}{R}\frac{\partial \Psi}{\partial R}+I\left(\Psi\right)\frac{\partial I}{\partial \Psi}=0.
\label{eq_pulsar}
\end{equation}
No analytical solution to the pulsar equation is known for a rotating dipole. However, an exact solution was found for a rotating monopole \citep{1973ApJ...180L.133M}. Some features of this solution are shared by the aligned dipole rotator, in particular in the wind zone beyond the light cylinder. Below we first describe the monopole solution and then discuss the dipole rotator.

\subsection{(Split) monopole: Michel's solution}

Michel's solution is given by (in spherical coordinates)
\begin{eqnarray}
B_{\rm r} &=& B_{\star}\left(\frac{r_{\star}}{r}\right)^2 \\
B_{\theta} &=& 0 \\
B_{\phi} &=& -B_{\star}\left(\frac{r_{\star}}{R_{\rm LC}}\right)\left(\frac{r_{\star}}{r}\right)\sin\theta \\
E_{\rm r} &=& 0 \\
E_{\rm \theta} &=& B_{\phi}\\
E_{\phi} &=& 0.
\label{eq_michel}
\end{eqnarray}
The solution is remarkably simple, as the poloidal magnetic field $B_P=B_r$ is unchanged from the normal monopole. Rotation creates the additional toroidal field $B_\phi=(V_\phi/c)B_r$ (where $V_\phi=\Omega r\sin\theta$ is the co-rotation velocity) and $E_\theta=B_\phi$. The corresponding current density is purely radial,
\begin{equation}
\mathbf{J}=\frac{c}{4\pi} \left(\mathbf{\nabla}\times\mathbf{B}\right) =-\frac{\bOm\cdot\bB}{2\pi}\,\mathbf{e}_r = c\rho\,\mathbf{e}_r,
\label{j_michel}
\end{equation}
where $\rho=\nabla\cdot\bE/4\pi$ is the
% corotation 
charge density. According to this force-free solution, the current can be carried by charge-separated plasma extracted from the star and moving with the speed of light. Since magnetic monopoles are forbidden by $\nabla\cdot\bB=0$, it is better to consider a ``split monopole,'' which is antisymmetric about the equatorial plane. Then the above solution describes one hemisphere, and the solution in the opposite hemisphere is obtained by the transformation $\mathbf{B}\rightarrow-\mathbf{B}$, $\mathbf{E} \rightarrow-\mathbf{E}$, and $\mathbf{J}\rightarrow-\mathbf{J}$. The reversal of the magnetic field across the equator implies the presence of an equatorial current sheet, which is not included in \Eq~(\ref{j_michel}). It carries the return current and ensures that the net current through the stellar surface vanishes, so the stellar charge does not grow.

\begin{figure}[]
\centering
\includegraphics[width=9cm]{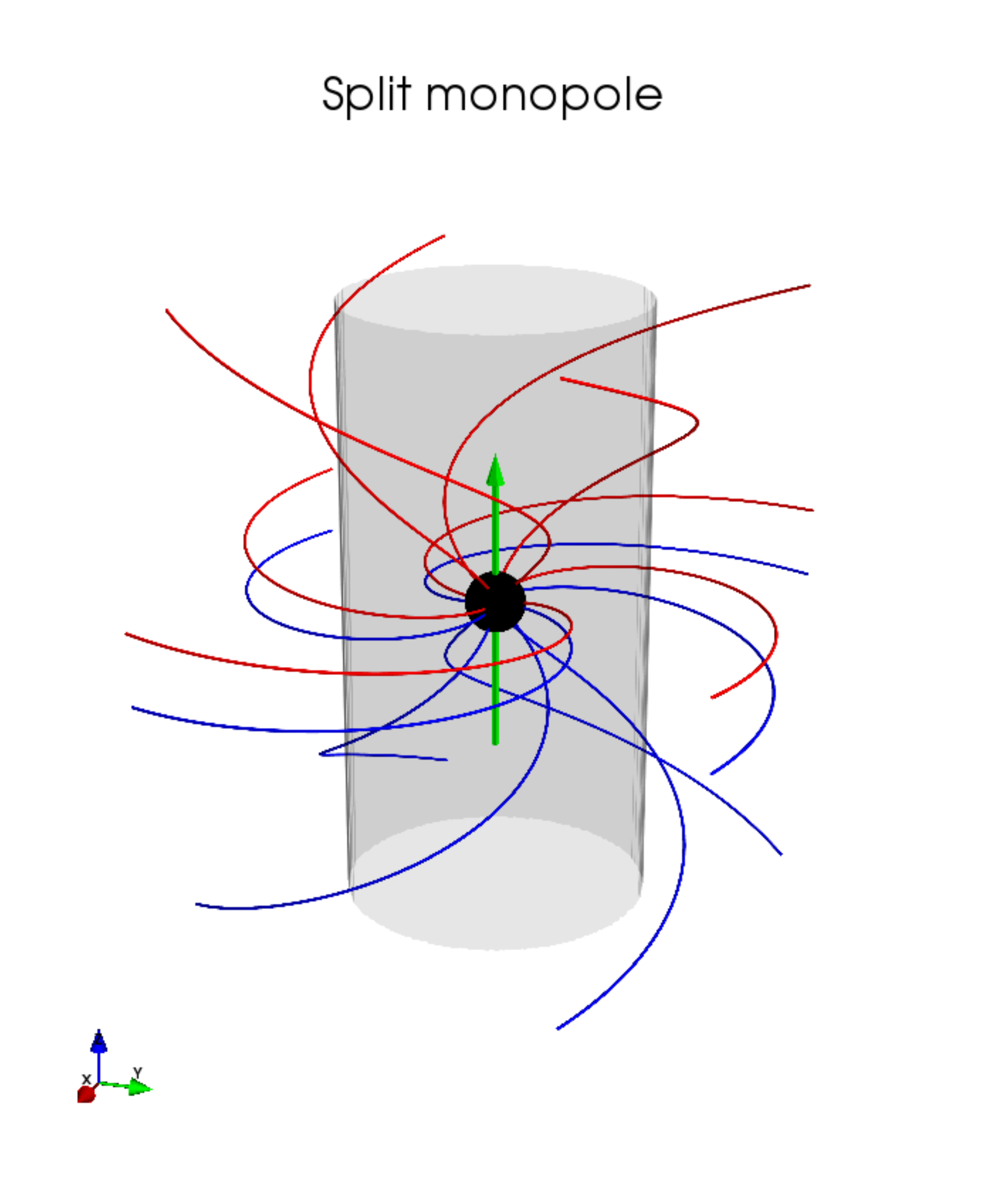}
\includegraphics[width=9cm]{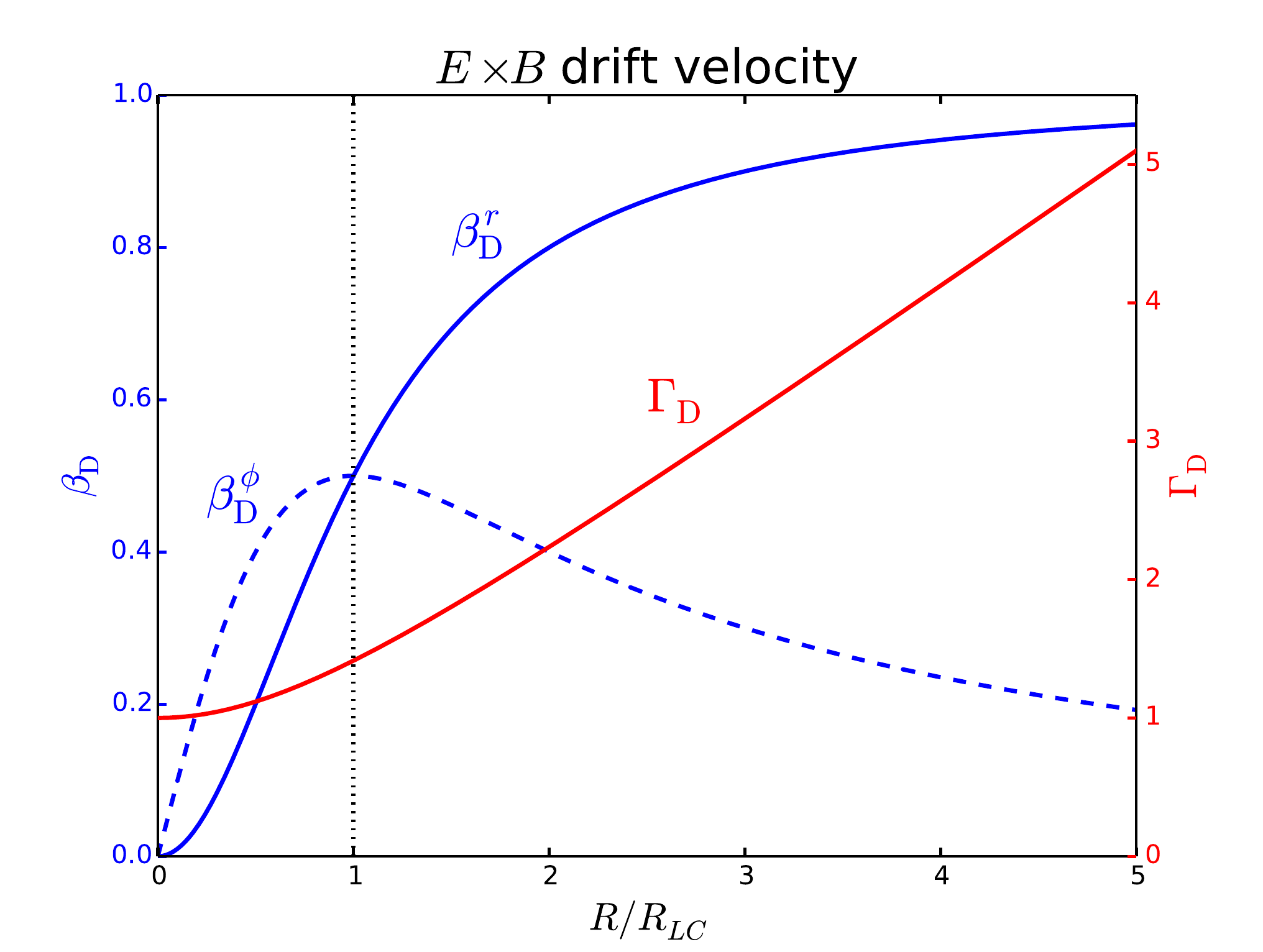}
\caption{Analytical solution of the aligned split monopole \citep{1973ApJ...180L.133M}. {\bf Top:} Magnetic field line structure winding up around the star (black sphere) due to the stellar rotation. The green arrow is the rotation axis of the star. Red lines represent outgoing field lines while blue lines show the incoming field lines. The field changes its polarity across the equator. The magnetosphere contained within the light-cylinder radius is shown in grey. {\bf Bottom:} $\mathbf{E}\times\mathbf{B}$ drift velocity (solid line: $\beta_{\rm D}^{\rm r}=V_{\rm D}^{\rm r}/c$, dashed line: $\beta_{\rm D}^{\phi}=V_{\rm D}^{\phi}/c$) and Lorentz factor, $\Gamma_{\rm D}=1/\sqrt{1-\beta_{\rm D}^2}=\sqrt{1+\left(R/R_{\rm LC}\right)^2}$, as a function of the cylindrical radius $R=r\sin\theta$.}

\label{fig_monopole}
\end{figure}

The relation $B_\phi/B_r=r\sin\theta/\RLC$ shows that the magnetic field lines are nearly radial close to the star ($r\sin\theta\ll R_{\rm LC}$) and start winding up around the pulsar with increasing distance (see Figure~\ref{fig_monopole}, top panel). At the light-cylinder, in the equatorial plane, the poloidal and toroidal magnetic components are equal, i.e. the field line makes a $45^{\rm o}$ angle with respect to the radial direction. Far outside the light cylinder, the magnetic field is almost purely toroidal. The Poynting flux $\mathbf{\Pi}=c\bE\times\bB/4\pi$ integrated over any sphere of radius $r>\rs$ determines the spindown power extracted from the star, 
\begin{equation}
L_{\rm mono}=\int \mathbf{\Pi}\cdot d\mathbf{S} =\frac{2cB^2_{\star}r^4_{\star}}{3R^2_{\rm LC}}.
\label{spindown_monopole}
\end{equation}
This result turns out also valid for any oblique split monopole, i.e. it does not depend on the inclination angle $\chi$ between $\bOm$ and the normal to the antisymmetry plane of $\bB$ \citep{1999A&A...349.1017B}. 

The drift velocity of the magnetic field lines is given by (Figure~\ref{fig_monopole}, bottom panel)
\begin{eqnarray}
\mathbf{V}_{\rm D} = c\frac{\mathbf{E}\times\mathbf{B}}{\mathbf{B}^2}= \frac{c~\mathbf{e_{\rm r}}}{1+\left(\frac{R_{\rm LC}}{R}\right)^2}+\frac{R\Omega~\mathbf{e_{\phi}}}{1+\left(\frac{R}{R_{\rm LC}}\right)^2}.
\label{eq_vdrift}
\end{eqnarray}
This velocity does not describe the true motion of the charge-separated plasma in the monopole magnetosphere. The force-free result $\bJ=c\rho\,{\mathbf e}_r$ implies that the charge-separated plasma everywhere flows radially with the speed of light, which is inconsistent with $\bV_{\rm D}$. The true plasma velocity can only be found beyond the realm of force-free models. The plasma is co-rotating with the star at the stellar surface and starts radial motion into the magnetosphere with a negligible $v_r$ and a very high charge density $\rho\gg J/c$. The Lorentz factor of the radial outflow is controlled by $E_\parallel\neq 0$ which requires a self-consistent calculation \citep{1974ApJ...192..713M}. The outflow Lorentz factor quickly grows with altitude and saturates at $\gamma\sim\sigma^{1/2}\cos\theta$, where $\sigma=e\Omega B_\star \rs^2/m_ec^3$. While $E_\parallel\neq 0$ is essential for the plasma acceleration, it is much smaller than $E_\theta$, $B_r$, $B_\phi$ when $\sigma\gg 1$. In this limit, the electromagnetic field around the star approaches the force-free model with $E_\parallel=0$.

The drift velocity profile given in Eq.~(\ref{eq_vdrift}) becomes relevant to plasma motion if the plasma has a high multiplicity of charges instead of being charge separated. In particular, the force-free pulsar wind of a high multiplicity (discussed below) approaches a split-monopole structure at $R\gg\RLC$. In this case, the plasma flies away with $\bV\approx\bV_{\rm D}$ until it reaches the fast magnetosonic point. It follows from Eq.~(\ref{eq_vdrift}) that the bulk Lorentz factor of the flow is given by
\begin{equation}
\Gamma_{\rm D}=\sqrt{1+\left(\frac{R}{R_{\rm LC}}\right)^2},
\end{equation}
so the pulsar wind accelerates linearly with cylindrical radius beyond the light cylinder \citep{1977MNRAS.180..125B, 2002ApJ...566..336C}.

\subsection{Dipole}

%%%%%%%%%%%%%%%%%%%%%%%
\begin{figure}[]
\centering
\includegraphics[width=10cm]{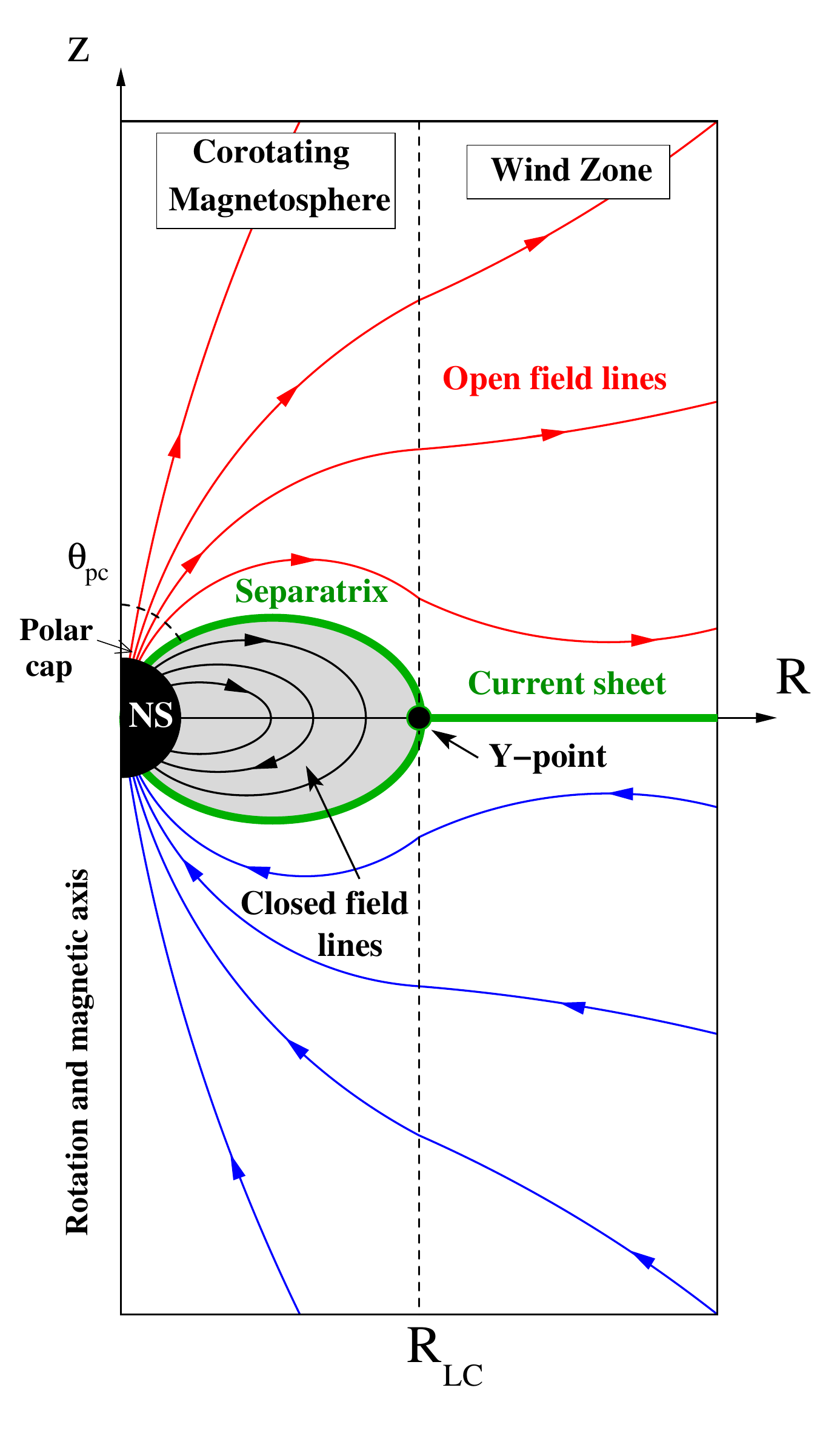}
\caption{Sketch of the ideal force-free magnetosphere of the aligned pulsar. The main elements are: (i) The closed field line region (grey, and black field lines) lying between the star surface and the light cylinder. This zone is dead and does not participate to the pulsar activity. (ii) The open field line region (red and blue field lines) extending beyond the light cylinder. The open field-line bundle carries the outflowing electric current, Poynting flux and the relativistic pulsar wind. (iii) The equatorial current sheet (green) between the opposite magnetic fluxes in the wind zone. It splits at the light cylinder into two separatrix current sheets that go around the closed zone, between the last open and the first closed field lines.}
\label{fig_ffe}
\end{figure}
%%%%%%%%%%%%%%%%%%%%%%%

The dipole magnetic flux emerging from the stellar surface is described by the flux function
\beq
\label{eq:dip}
\Psi(\rs,\theta)=\frac{\mu\sin^2\theta}{\rs}.
\eeq
As in the monopole case, rotation of the star $\bOm$ distorts the external dipole field. The force-free magnetosphere of the aligned rotator ($\bOm\parallel\bmu$) is described by the poloidal flux function $\Psi(r,\theta)$ that obeys the pulsar equation (\Eq~\ref{eq_pulsar}) with the boundary condition \Eq~(\ref{eq:dip}). This equation was solved numerically by \citet{1999ApJ...511..351C} (see also \citealt{2004MNRAS.349..213G, 2005PhRvL..94b1101G, 2005A&A...442..579C, 2006MNRAS.368.1055T}). The solution gives the poloidal magnetic field $\bB_P$ (which determines the charge density $\rho_{\rm GJ}=-\bOm\cdot\bB_P/2\pi c$) and the self-consistent poloidal current $\bJ_P$ (which determines $B_\phi$).

The force-free magnetosphere possesses the following three basic features depicted schematically in Figure~\ref{fig_ffe}: (i) an open zone with $\bJ_P\neq 0$ and $B_\phi\neq 0$, resembling the monopole solution, especially far outside the light cylinder, (ii) a closed zone with $\bJ_P=0$ and $B_\phi=0$, and (iii) a Y-shaped current sheet. The current sheet supports the jump in $B_\phi$ between the open and closed zones inside the light cylinder (the separatrix) and the jump $\bB\rightarrow -\bB$ across the equatorial plane in the wind zone. The current sheet is positively charged outside the light cylinder and negatively charged inside, as predicted by \citet{1990SvAL...16...16L}. The presence of the closed zone and the Y-point in the magnetic field topology is an essential difference from the monopole magnetosphere. The closed magnetic field lines are not much different from a normal dipole configuration. However, the closed zone is not vacuum --- it is filled with charge density $\rho_{\rm GJ}$, similar to the corotating part of the torus in the electrosphere solution. It co-rotates with the star, creating current $J_\phi=V_\phi\rho_{\rm GJ}$ which increases the effective dipole moment of the star, slightly inflating the poloidal magnetic field lines.

The field lines in the open zone would be closed if there were no rotation; rotation pushes them to infinity. This effect may be thought of as a result of the effective inertial mass of the magnetic field $B^2/8\pi c^2$ (\citealt{1973ApJ...180..207M}), as there is no plasma inertia in the force-free magnetosphere. Like the monopole case, the field outside the light cylinder becomes increasingly dominated by the toroidal component, and the field energy flows out with nearly speed of light. The amount of open flux $\Psi_{\rm pc}$ may be estimated approximating the closed zone by the dipole field. Then the last closed field line crosses the stellar surface at the polar angle $\theta_{\rm pc}$ given by
\begin{equation}
\sin^2\theta_{\rm pc}=\frac{r_{\star}}{R_{\rm LC}}, 
\qquad \Psi_{\rm pc}=\frac{\mu}{R_{\rm LC}}.
\label{eq_thetapc}
\end{equation}
The angle $\theta_{\rm pc}$ defines the ``polar cap'' --- the footprint of the open field-line bundle on the star, carrying magnetic flux $\Psi_{\rm pc}$. Its accurate value is slightly larger, because of the poloidal inflation of magnetic field lines caused by $J_\phi>0$. The open field-line bundle carries the polar-cap current and the Poynting flux into the wind zone. This is the active part of the magnetosphere which spins down the star. 

The spindown power may be estimated using the fact that $B_\phi\sim B_P$ and $E_\theta\sim B_P$ near the light cylinder (similar to the monopole magnetosphere). The net radial Poynting flux may be estimated as $L\approx 4\pi\RLC^2 E_\theta B_\phi/4\pi c$, which gives
\begin{eqnarray}
L\approx  \frac{\mu^2\Omega^4}{c^3}
\qquad {\rm (aligned~rotator).}
\label{eq_aligned}
\end{eqnarray}
This simple expression turns out quite accurate, within a numerical factor close to unity.

\subsection{Time-dependent force-free simulations}

In the standard force-free solution for the aligned rotator, the last closed field line touches the light-cylinder radius, i.e. the Y-point is at $r=\RLC$. \citet{2006MNRAS.368.1055T} pointed out that this is not required by the pulsar equation, and in fact there is a continuous sequence of steady solutions with the Y-point located at $r\leq \RLC$. This degeneracy is resolved by direct {\it time-dependent} simulations of the force-free magnetosphere, which show how the steady state is established \citep{2006ApJ...648L..51S, 2006MNRAS.368L..30M, 2012MNRAS.423.1416P}.

The force-free electrodynamic equations can be formulated exclusively in terms of $\bE$ and $\bB$ with no reference to the plasma. Combining \Eq~(\ref{eq_ff}) with $\partial \left(\mathbf{E}\cdot\mathbf{B}\right)/\partial t=0$, one obtains  \citep{2002luml.conf..381B}
\begin{equation}
\mathbf{J}=c\rho\left(\frac{\mathbf{E}\times\mathbf{B}}{\mathbf{B}^2}\right)+\frac{c}{4\pi}\left(\mathbf{B}\cdot\nabla\times\mathbf{B}-\mathbf{E}\cdot\nabla\times\mathbf{E}\right)\frac{\mathbf{B}}{\mathbf{B}^2}.
\label{eq:J}
\end{equation}
Using $\rho=\nabla\cdot\bE/4\pi$ and substituting \Eq~(\ref{eq:J}) into the Maxwell equation $\partial\bE/\partial t=\nabla\times\bB-(4\pi/c)\bJ$ one can express $\partial\bE/\partial t$ in terms of $\bE$, $\bB$, and their spatial derivatives. Together with $\partial\bB/\partial t=-c\nabla\times\bE$, this gives a closed dynamical system of field equations that can be evolved in time. For example, one can start with a non-rotating star with a vacuum dipole magnetosphere, then spin it up to a desired $\bOm$, and let it relax to the quasi-steady state. For the aligned rotator, these simulations reproduced the steady-state solution with the Y-point located close to the light cylinder.
  
  \begin{figure}[]
\centering
\includegraphics[width=4.2cm]{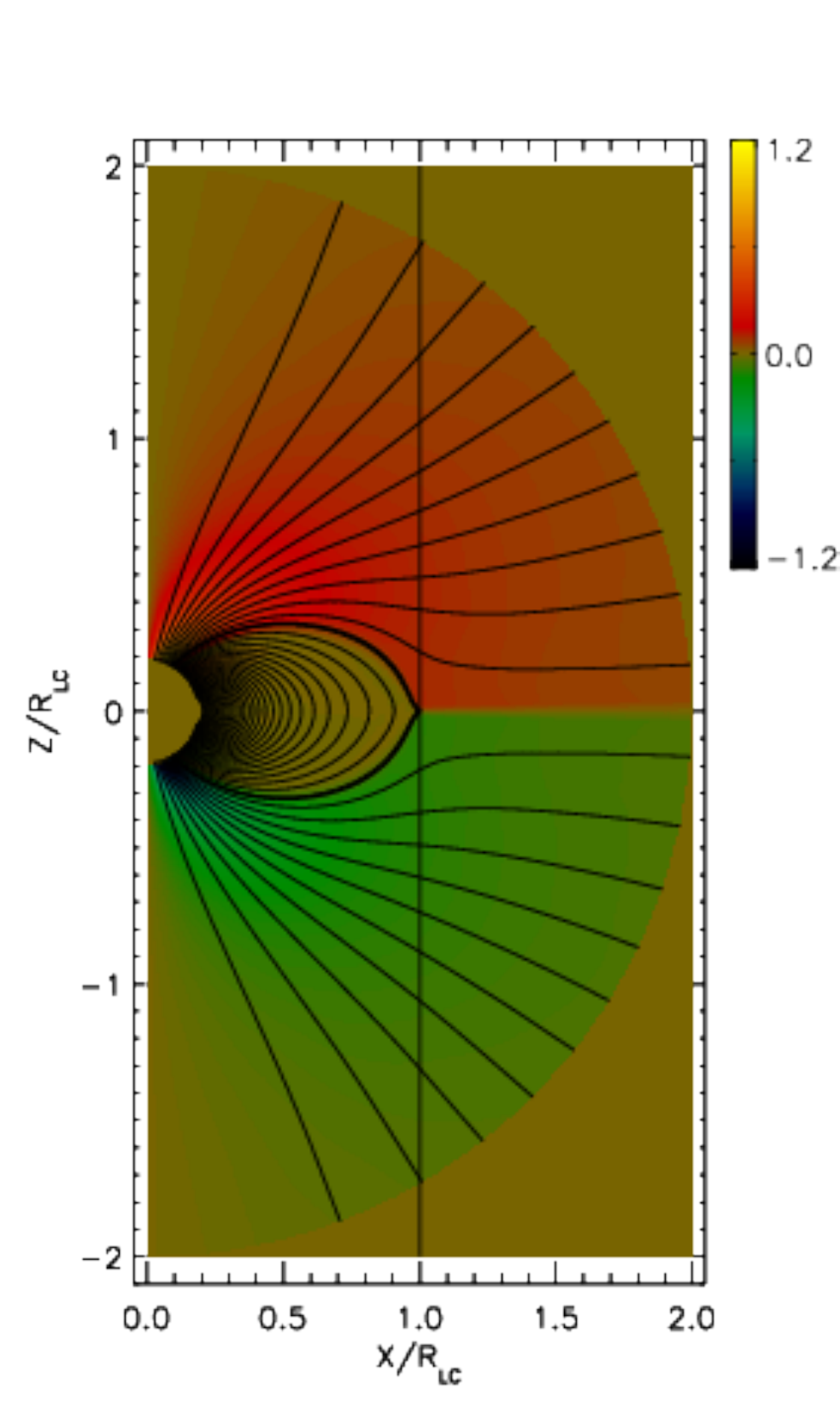}
\includegraphics[width=7.5cm]{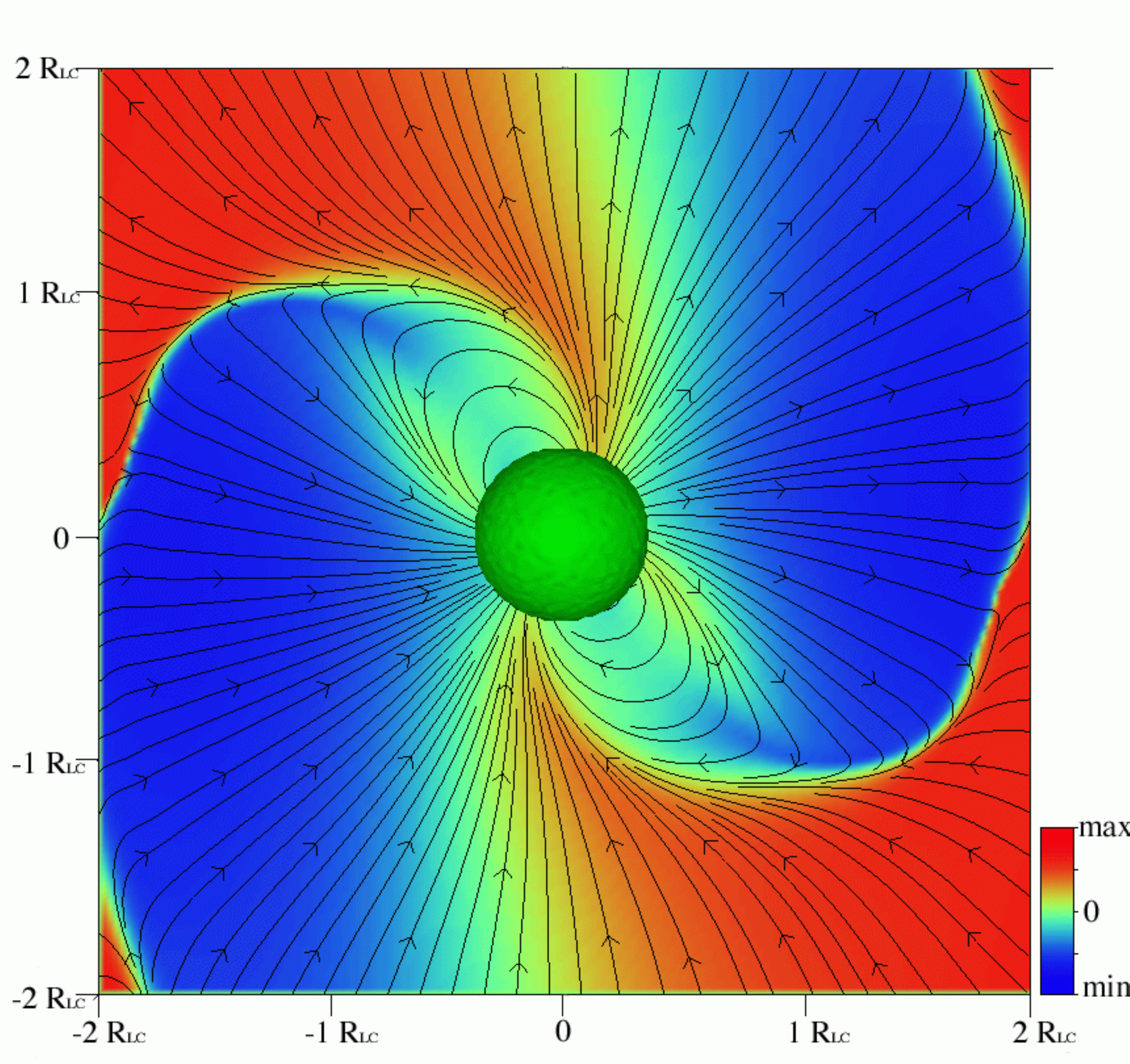}
\caption{Snapshots of time-dependent force-free simulations of the aligned (left) and oblique (right) rotators (from \citealt{2006ApJ...648L..51S}). The oblique rotator magnetosphere is shown in the $\mathbf{\Omega}-\boldsymbol{\mu}$ plane; the inclination angle is $\chi=60^{\rm o}$. Solid lines represent magnetic field lines, and color shows the strength of the magnetic field component perpendicular to the plane of the figure (the toroidal field in the aligned rotator case).}
\label{fig_ffsim}
\end{figure}

%%%%%%%%%%%%%%%%%%%%%%%%%%%%%
\begin{figure}[]
\centering
\includegraphics[width=11cm]{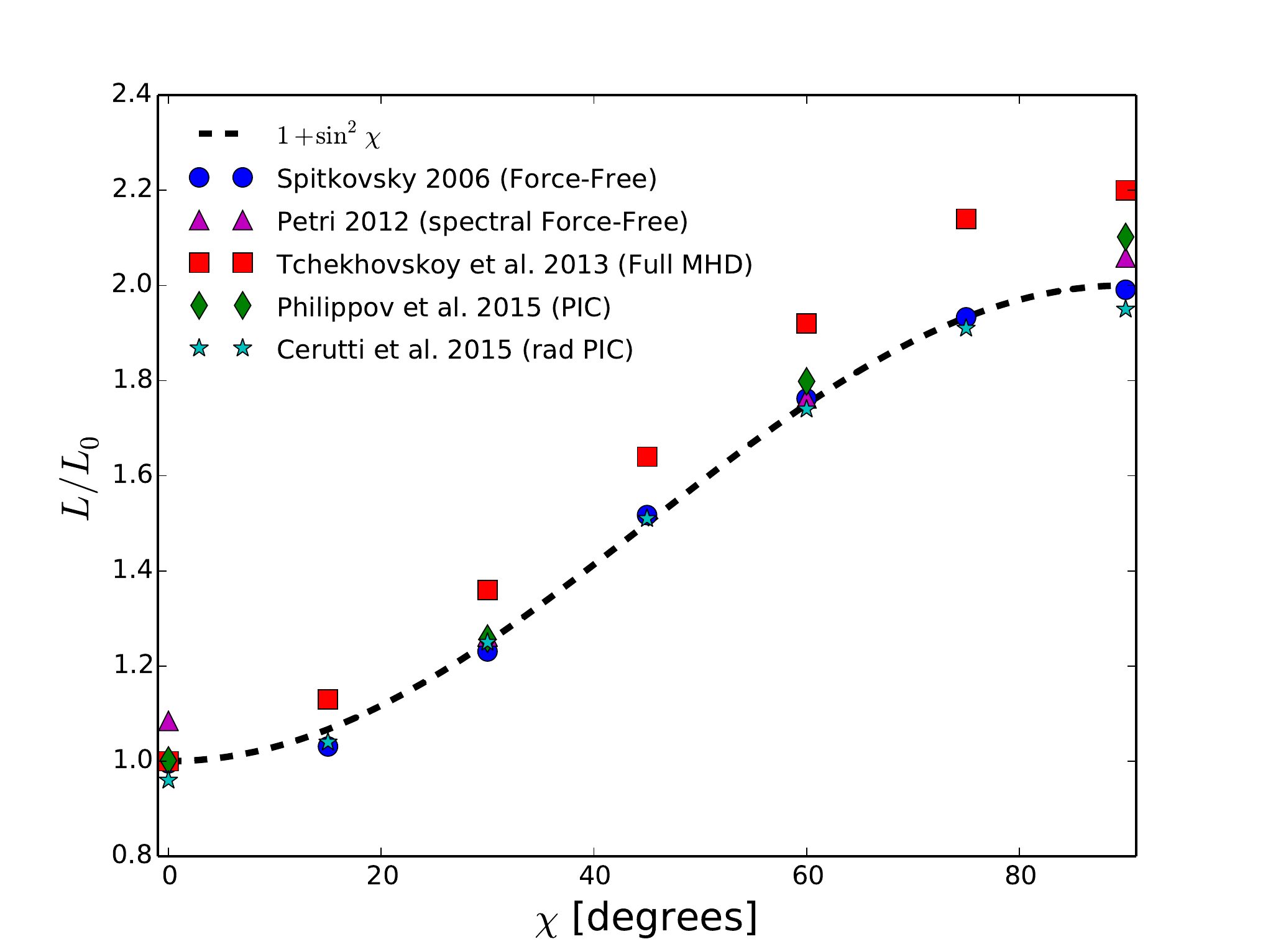}
\caption{Compilation of the spindown power for the oblique plasma-filled magnetosphere reported in the literature in different numerical models: force-free \citep{2006ApJ...648L..51S}, spectral force-free \citep{2012MNRAS.424..605P}, full MHD \citep{2013MNRAS.435L...1T}, PIC \citep{2015ApJ...801L..19P}, and radiative PIC \citep{2016MNRAS.457.2401C}. The spindown power is normalized by $L_0=\mu^2\Omega^4/c^3$. The black dashed line shows the approximate analytical fit $L\approx L_0\left(1+\sin^2\chi\right)$.}
\label{fig_spindown}
\end{figure}
%%%%%%%%%%%%%%%%%%%%%%%%%%%%%

Three-dimensional time dependent simulations were then performed for the oblique rotator (right panel in Figure~\ref{fig_ffsim}, \citealt{2006ApJ...648L..51S, 2009A&A...496..495K, 2012ApJ...749....2K, 2012MNRAS.424..605P}). These simulations showed how the spindown power depends on the angle $\chi$ between $\bOm$ and $\bmu$ (Figure~\ref{fig_spindown}). The result is well approximated  by a simple formula proposed by \citet{2006ApJ...648L..51S}, 
\begin{equation}
L\approx\frac{\mu^2\Omega^4}{c^3}\left(1+\sin^2\chi\right).
\label{spindown_oblique}
\end{equation}
In contrast to the oblique split monopole, the dipole spindown power depends on $\chi$. The term proportional to $\sin^2\chi$ is similar to the vacuum model, except for a different numerical prefactor (1 instead of 2/3). Note also that $L$ does not vanish in the aligned case $\chi=0$. The origin of the variation $L(\chi)$ was discussed by \citet{2016MNRAS.457.3384T}. Part of it comes from the increasing open magnetic flux $\Psi_{\rm pc}(\chi)$ (explaining 40\% of the increase in $L$), and the remaining part is caused by the increasing concentration of the open magnetic flux toward the equator. Apart from the moderate dependence $L(\chi)$, the wind from the dipole rotator is similar to that from a split monopole with equal open magnetic flux.

Force-free simulations have also verified the existence of the current sheet in the dipole magnetosphere. Its presence is important as it carries most of the return current, which makes the net current through each polar cap equal to zero. In the wind zone, the current sheet separates the two opposite magnetic fluxes. For the aligned rotator, the current sheet is flat outside the light cylinder and lies in the equatorial plane. For an oblique rotator, the current sheet outside the light cylinder has the shape of a ballerina skirt (``striped wind'', a relativistic analog of the heliospheric current sheet). It oscillates in $\theta$ around $\pi/2$ with the amplitude $\chi$ and a wavelength of $2\pi R_{\rm LC}$.

A key feature of the equatorial current sheet is that it is the place where the force-free approximation certainly breaks down. The magnetic field flips across the sheet, i.e. it must pass through $B=0$, making the plasma unmagnetized (infinite Larmor radius) and violating the force-free condition $B^2/8\pi \gg U_{\rm pl}$ where $U_{\rm pl}$ is the plasma energy density including its rest-mass energy. A complete model of the current sheet should include the plasma inertia and pressure, a finite resistivity and localized dissipation, which promotes magnetic reconnection. In numerical force-free models, the singular behavior of the current sheet outside the light cylinder is regularized by introducing artificial resistivity \citep{2012ApJ...746...60L, 2012MNRAS.423.1416P, 2012ApJ...749....2K}. Alternatively, the full MHD approach can be employed to model the plasma inertia and pressure in the sheet \citep{2006MNRAS.367...19K, 2013MNRAS.435L...1T}.

%###########################################################

\section{Toward a realistic magnetosphere: pair creation, particle acceleration and radiation}\label{sect_pic}

Observations show that the spindown power is partially dissipated in the magnetosphere, resulting in bright X-ray and gamma-ray emission. Radiation is produced by relativistic particles and a self-consistent model is needed that would demonstrate how and where the particles are created and accelerated. A traditional approach to pulsar modeling was based on guessing the location of electric gaps inside the light cylinder --- places where a strong electric field $E_\parallel$ accelerates particles to relativistic energies. Various gaps were proposed: polar-cap gap \citep{1971ApJ...164..529S, 1975ApJ...196...51R}, slot gap \citep{1983ApJ...266..215A, 2004ApJ...606.1143M}, and outer gap \citep{1986ApJ...300..500C}. The approach to this problem significantly evolved in the last decade, as described below.

\subsection{Hints from the force-free model}

The force-free model shows one obvious candidate for strong emission --- the equatorial current sheet outside the light cylinder, as anticipated in some earlier work \citep{1994MNRAS.271..621M, 1996A&A...311..172L}. In addition, the force-free model provides guidance to possible locations of the particle accelerator inside the light cylinder, since it shows the current density $\bJ$ and charge density $\rho$ desired by the magnetosphere. Using the dimensionless parameter $\alpha=J/c\rho$, one can formulate a simple necessary condition for particle acceleration \citep{2008ApJ...683L..41B}:
\beq
\label{eq:alpha}
\alpha^{-1} \leq 1, \qquad \alpha\equiv\frac{J}{c\rho},
\eeq
which corresponds to $\alpha\geq 1$ or $\alpha<0$. If the opposite condition $0\leq\alpha<1$ is satisfied along the magnetic field line, the charge-separated flow extracted from the star is pushed to velocity $\bv=\bJ/\rho$ and provides the desired $J$ and $\rho$ with insignificant acceleration. The condition $0\leq\alpha<1$ corresponds to $0\leq v<c$, making the velocity $v=J/\rho$ accessible to the flow, which is sufficient for avoiding gaps. In contrast, if $\alpha^{-1}\geq 1$, the charge-separated outflow either fails to conduct the current (and then $E_\parallel$ is generated by the induction effect) or fails to sustain $\rho$. In this case, the magnetosphere may approach the desired configuration only if the field lines are loaded with $e^\pm$ pairs created in the magnetosphere --- otherwise a gap with a strong unscreened $E_\parallel$ must form.
   
This criterion can be applied to the polar cap of the aligned dipole rotator using $\alpha$ calculated for the force-model by \citet{2006MNRAS.368.1055T}, see also \citet{2012MNRAS.423.1416P}. On the polar-cap surface, $\alpha(\theta)$ decreases from unity at $\theta=0$ and becomes negative near the edge of the polar cap $\theta_{\rm pc}$. The return current sheet at the edge also has $\alpha<0$ (recall that the force-free magnetosphere requires the current sheet inside the light cylinder both to be negatively charged and to conduct $J_r>0$).
   
The criterion in \Eq~(\ref{eq:alpha}) then says that the polar-cap accelerator is suppressed in most of the polar-cap interior --- here the extracted charge-separated flow with $v=\alpha c$ screens $E_\parallel$. More exactly, a moderate $E_\parallel$ is sustained to move the plasma from the surface, and the flow velocity $v$ oscillates around $\alpha c$ \citep{1985MNRAS.217..443M, 2008ApJ...683L..41B}. The resulting oscillating flow has the maximum dimensionless momentum $p=\gamma\beta$ given by
\beq
\label{eq:pmax}
p_{\max}=\frac{2\alpha}{1-\alpha^2}   \qquad (0<\alpha<1).
\eeq
This result was confirmed by one-dimensional particle-in-cell (PIC) simulations of the charge-separated flow above the polar cap \citep{2013ApJ...762...76C, 2013MNRAS.429...20T}.

The fact that $\alpha<0$ near the edge of the polar cap implies strong particle acceleration, in particular in the current sheet at the boundary with the closed magnetosphere. One can show that $e^\pm$ creation in the accelerator with $\alpha^{-1}\leq 1$ must be time-dependent and follow a limit-cycle pattern \citep{2008ApJ...683L..41B}. The cyclic discharge is demonstrated by one-dimensional PIC simulations that track $e^\pm$ creation \citep{2013MNRAS.429...20T}. This is a rather general property of $e^\pm$ discharge; similar one-dimensional PIC simulations of pair creation in the twisted closed magnetospheres of magnetars (which have $\alpha\gg 1$) also exhibit quasi-periodic behavior \citep{2007ApJ...657..967B}.

\subsection{Global numerical experiments: particle-in-cell (PIC) simulations}
 
The full problem of gap formation and particle acceleration is a global problem, because particles and pair-creating photons can flow through the magnetosphere. For instance, above the polar cap, the condition $\alpha^{-1}\leq 1$ is necessary but may not be sufficient for gap formation, as $e^\pm$ pairs may be supplied to the polar cap (and screen $E_\parallel$) from an accelerator located elsewhere in the magnetosphere. Simulations of the plasma-filled magnetosphere developed in recent years  \citep{2014ApJ...785L..33P, 2014ApJ...795L..22C, 2015ApJ...801L..19P, 2015ApJ...815L..19P, 2015MNRAS.448..606C, 2016MNRAS.457.2401C, 2015MNRAS.449.2759B, 2015NewA...36...37B} provided an opportunity to study this problem from first principles.
   
In the full global problem, it is not known in advance where the $e^\pm$ discharge will occur, and the problem requires a self-consistent calculation of the electromagnetic field, plasma dynamics and $e^\pm$ creation in a region that extends from the stellar surface to $r\gg \RLC$. Self-consistent simulations of $e^\pm$ discharge showing how the magnetosphere is filled with plasma were first performed for an aligned rotator \citep{2014ApJ...795L..22C} and then for an inclined rotator \citep{2015ApJ...801L..19P}. They demonstrated that the main particle accelerator in pulsars is the Y-shaped current sheet.

The basic technique for such simulations is the particle-in-cell (PIC) method \citep{1991ppcs.book.....B}. The technique was originally developed in the 1960s to study kinetic problems in plasma physics (e.g. \citealt{1962PhFl....5..445D}). It consists of describing a collisionless plasma from its fundamental constituents --- individual charged particles (``macroparticles'') and an electromagnetic field. Each macroparticle represents a large number of real particles which would be following the exact same path in phase space. From the positions and velocities of each macroparticle, the charge and current densities are deposited on the numerical grid. Then, the full time-dependent Maxwell's equations are integrated on a staggered mesh to ensure second order accuracy in both space and time \citep{1966ITAP...14..302Y}. All these steps are done within each time step, so that the motion of particles and the evolution of the electromagnetic fields are captured self-consistently.

In contrast to other applications (shocks, reconnection), PIC simulations of pulsars must be global, as the problem is intrinsically non-local --- what is happening at the light-cylinder affects the activity at the polar cap and vice-versa. Such simulations are challenging because the separation between the macroscopic ($r_{\star}$, $R_{\rm LC}$) and microscopic (plasma skindepth) scales is huge in real pulsars. In practice, simulations are limited to a few orders of magnitude in scale separation, which is sufficient to capture basic physics of pulsars, such as the mechanism of gap formation and production of gamma-rays.

\begin{figure}[t]
%  \centering
\hspace*{-3mm}
\includegraphics[width=12.6cm]{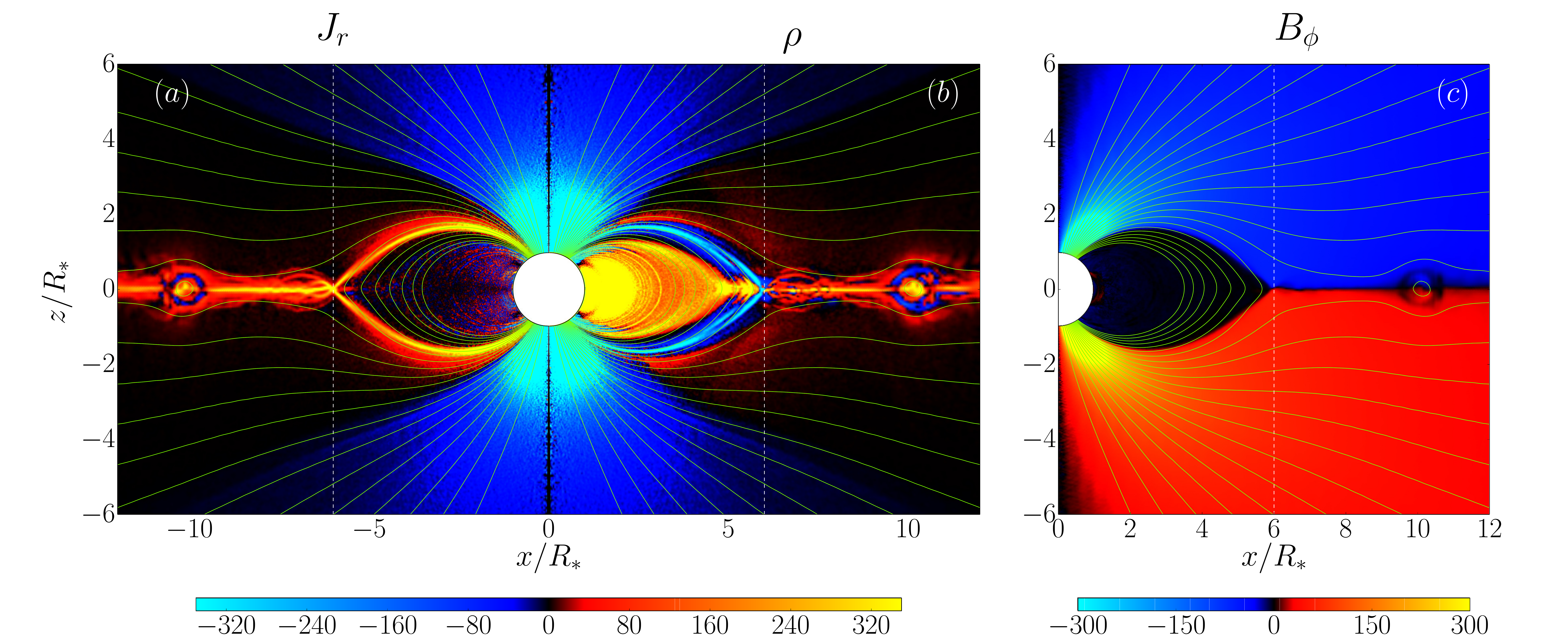} 
\hspace*{-3.7mm}
\includegraphics[width=8.88cm]{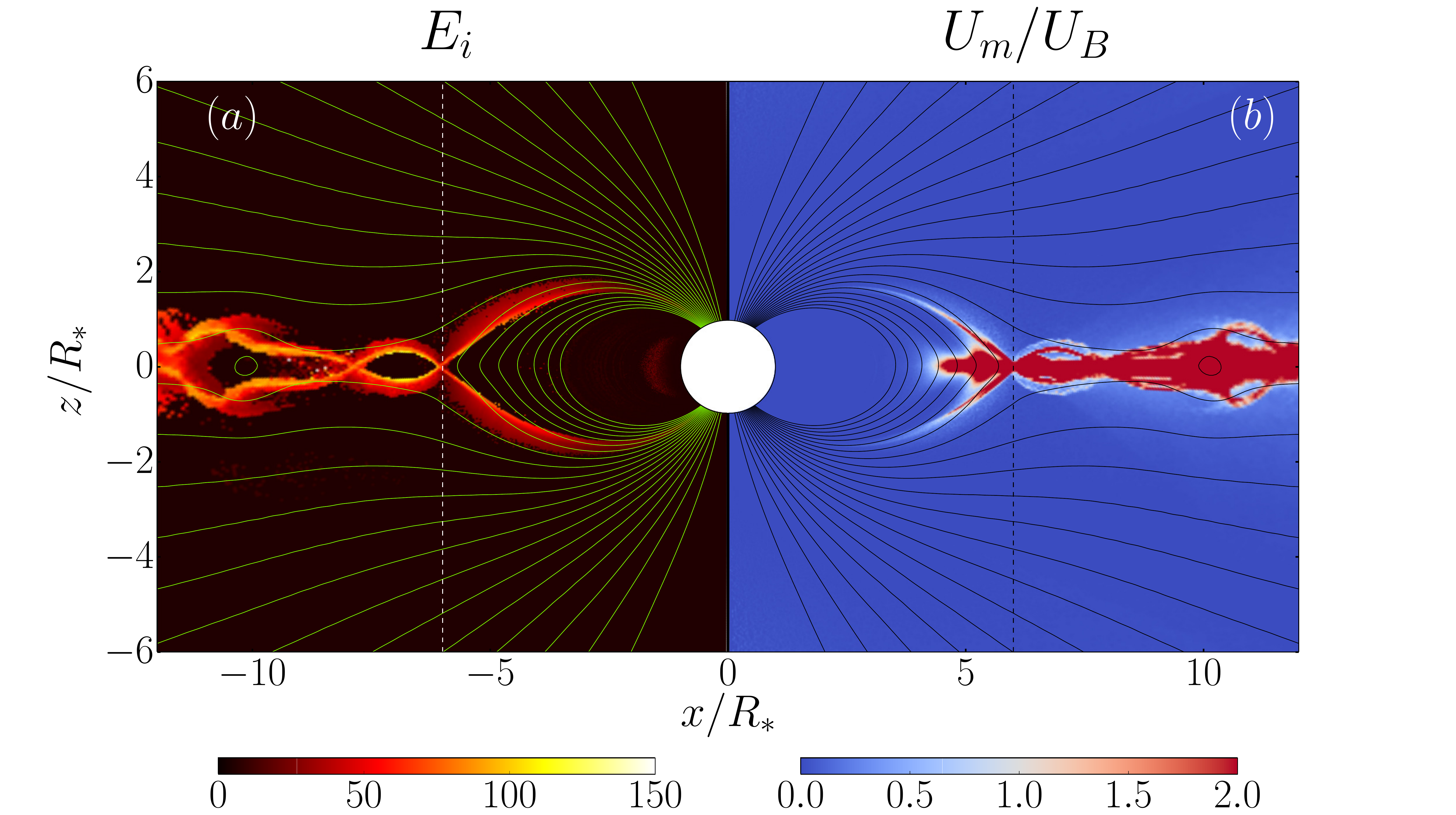}
\caption{Poloidal cross section of the magnetosphere of the aligned rotator (from \citealt{2014ApJ...795L..22C}). Vertical dashed line shows the light cylinder $\RLC=c/\Omega$. Green curves show the poloidal magnetic field lines. {\bf Top:} (a) Radial component of the electric current density $J_r$. (b) Net charge density $\rho$. (c) Toroidal component of the magnetic field $B_\phi$. Units: distance is measured in $\rs$, charge density in $m_e c^2/4\pi e\rs^2$, current in $m_{\rm e} c^3/4\pi e\rs^2$, and field in $m_{\rm e}c^2/e\rs$. {\bf Bottom:} (a) Average ion energy in units of $m_{\rm e}c^2$  (the ion rest mass was re-scaled to $5m_{\rm e}$ in the simulation). One can see the acceleration of ions in the gap, their ejection through the Y-point, and gyration in the equatorial current sheet. (b) Ratio of matter energy density $U_m$ to magnetic energy density $U_B=B^2/8\pi$.}
\label{fig_pic2D}
\end{figure}

Apart from the numerical PIC aspects, a major challenge in pulsar simulations is the self-consistent implementation of $e^\pm$ creation. It was first implemented in an axisymmetric PIC simulation by \citet{2014ApJ...795L..22C}. Since it was not known in advance where pairs will be created, the simulation started with a ``clean'' initial condition --- a non-rotating vacuum dipole, which was gradually spun up to a desired $\Omega$. It was observed how the rotation-induced electric field extracted particles from the stellar surface and accelerated them to high Lorentz factors leading to curvature emission, and how the curvature photons converted to $e^\pm$ pairs. This numerical experiment demonstrated the mechanism of filling the magnetosphere with plasma, with the following results:
\begin{itemize}
\item 
There are two types of pulsar magnetospheres, depending on the physical process available to create $e^\pm$ pairs. If $e^\pm$ are created only in a strong magnetic field (due to gamma-ray absorption by the field, \citealt{1966RvMP...38..626E}) then pair production is limited to $r\ll\RLC$. \citet{2014ApJ...795L..22C} called it ``Type II'' rotator. In this case, the aligned rotator was found to evolve to a nearly ``dead'' state, resembling the electrosphere described in Section~\ref{sect_elec}. It is qualitatively different from Type I rotators where $e^\pm$ pairs can be created at $r\sim\RLC$ through photon-photon collisions.
\item 
Type I rotator {\it self-organizes to produce copious pairs inside and around the Y-shaped current sheet.} It evolves into a quasi-steady magnetic configuration close to the force-free solution (Figure~\ref{fig_pic2D}).
\item
The current sheet inside the light cylinder (the separatrix) develops a time-dependent gap stretched along the closed zone boundary: {\it the separatrix gap.} This accelerator enables pair creation required to sustain the current sheet and the open magnetic flux. The gap is qualitatively different from the ``slot gap'' and ``outer gap'' proposed in earlier works. The earlier models assumed that the charge density ``desired'' by the magnetosphere is $\rhoGJ$. In contrast, the charge density in the current sheet greatly exceeds $\rhoGJ$ (in the ideal force-free model it would be infinite). The accelerating voltage develops because the large $\rho$ and $J$ (with $\alpha<0$) cannot be sustained without a high rate of pair creation.
\end{itemize}
      
%%%%%%%%%%%%%%%%%%%%%%%%%%%%%%%%%%%%%
%%%%%%%%%%%%%%%%%%%%%%%%%%%%%%%%%%%%%
      
The global PIC simulations also show how the plasma is accelerated and ejected through the Y-point into the equatorial current sheet (Figure~\ref{fig_pic2D}). Plasma outflows along the equatorial plane outside $\RLC$ and the Y-point resembles a nozzle formed by the open magnetic fluxes of opposite polarity. Two plasma streams come to the Y-point along the boundary of the closed zone and exchange their opposite $\theta$-momenta. Their collimation outside the light cylinder is achieved through gyration in the (predominantly toroidal) magnetic field, which communicates the $\theta$-momentum from one stream to the other.   

\begin{figure}[]
\centering
 \includegraphics[width=12.0cm]{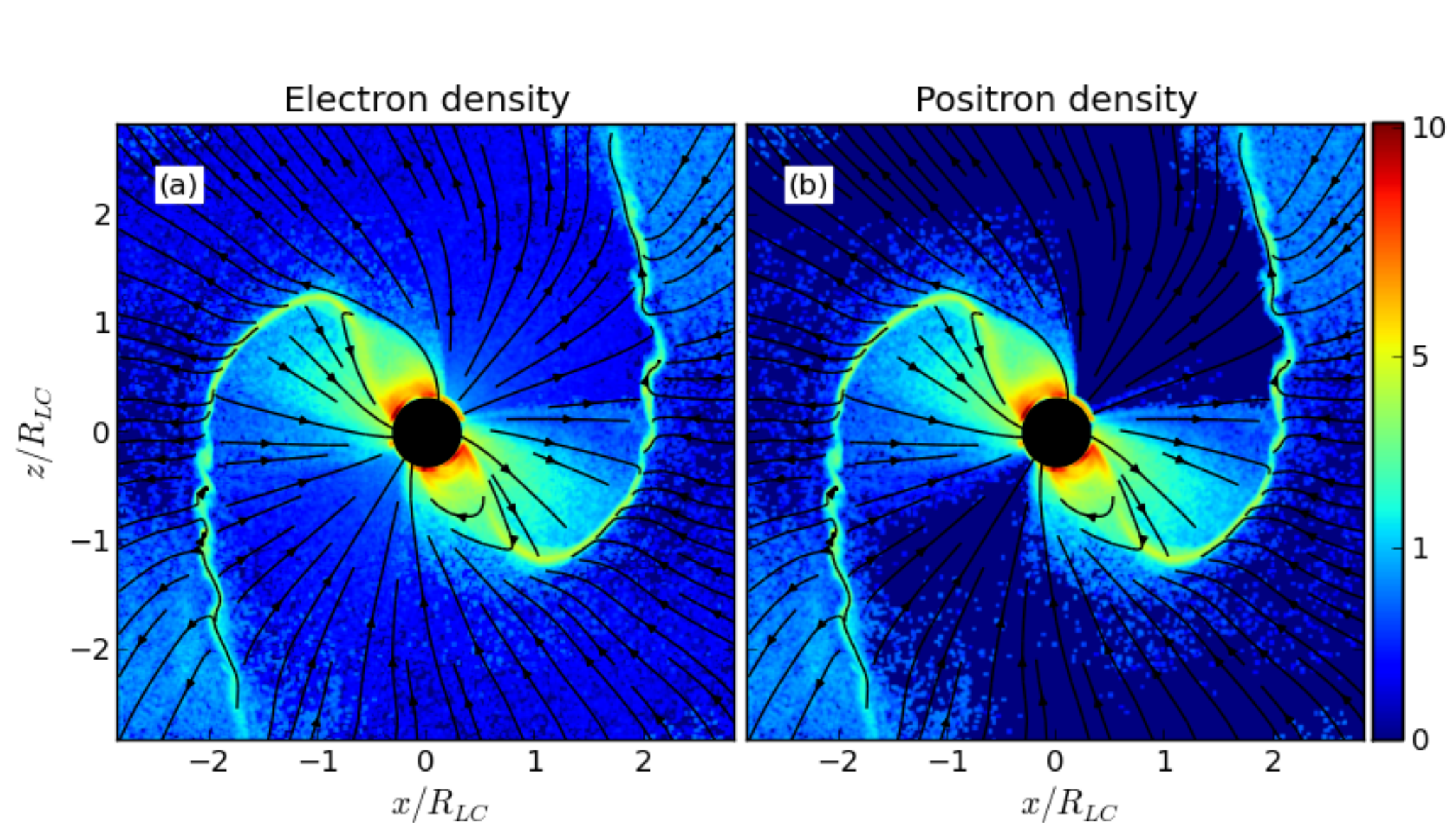}
\caption{Electron (left) and positron (right) densities in a 3D PIC simulation of the oblique rotator with $\chi=60^{\rm o}$ (from \citealt{2015ApJ...801L..19P}). The magnetosphere is viewed in the $\mathbf{\Omega}-\boldsymbol{\mu}$ plane; solid curves show the magnetic field lines. Particle acceleration and $e^\pm$ creation occurs in part of the polar cap as well as in the current sheet.}
\label{fig_pic}
\end{figure}

The 3D simulations of inclined rotators \citep{2015ApJ...801L..19P} show a similar mechanism of sustaining the wobbling current sheet predicted by the 3D force-free simulations of \citet{2006ApJ...648L..51S}. A snapshot of the inclined rotator with $\chi=60^{\rm o}$ is shown in Figure~\ref{fig_pic}.

%%%%%%%%%%%%%%%%%%%%%%
%%%%%%%%%%%%%%%%%%%%%%

%%%%%%%%%%%%%%%%%%%%%%%%%%
\begin{figure}[]
\centering
\includegraphics[width=11cm]{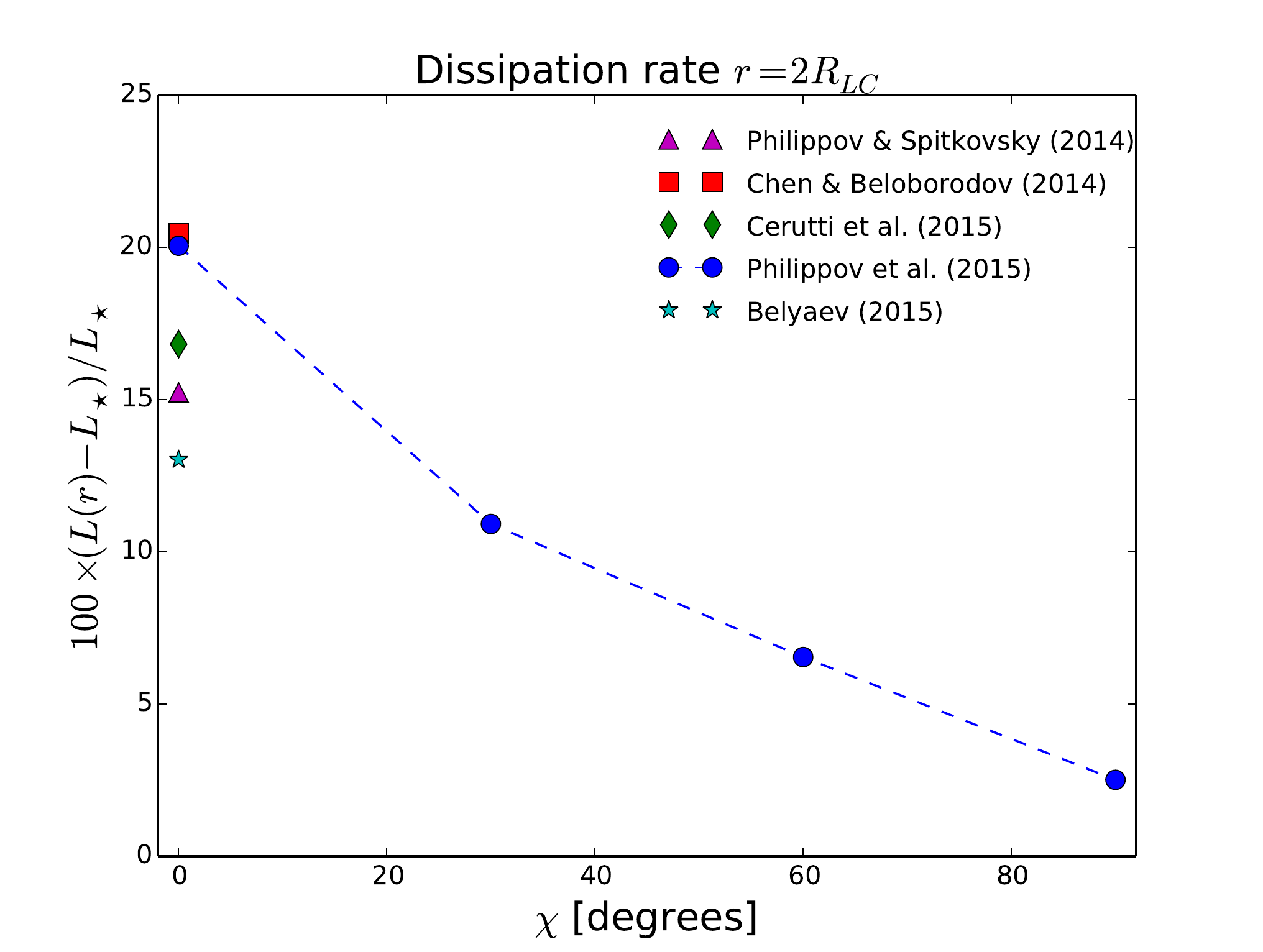}
\includegraphics[width=11cm]{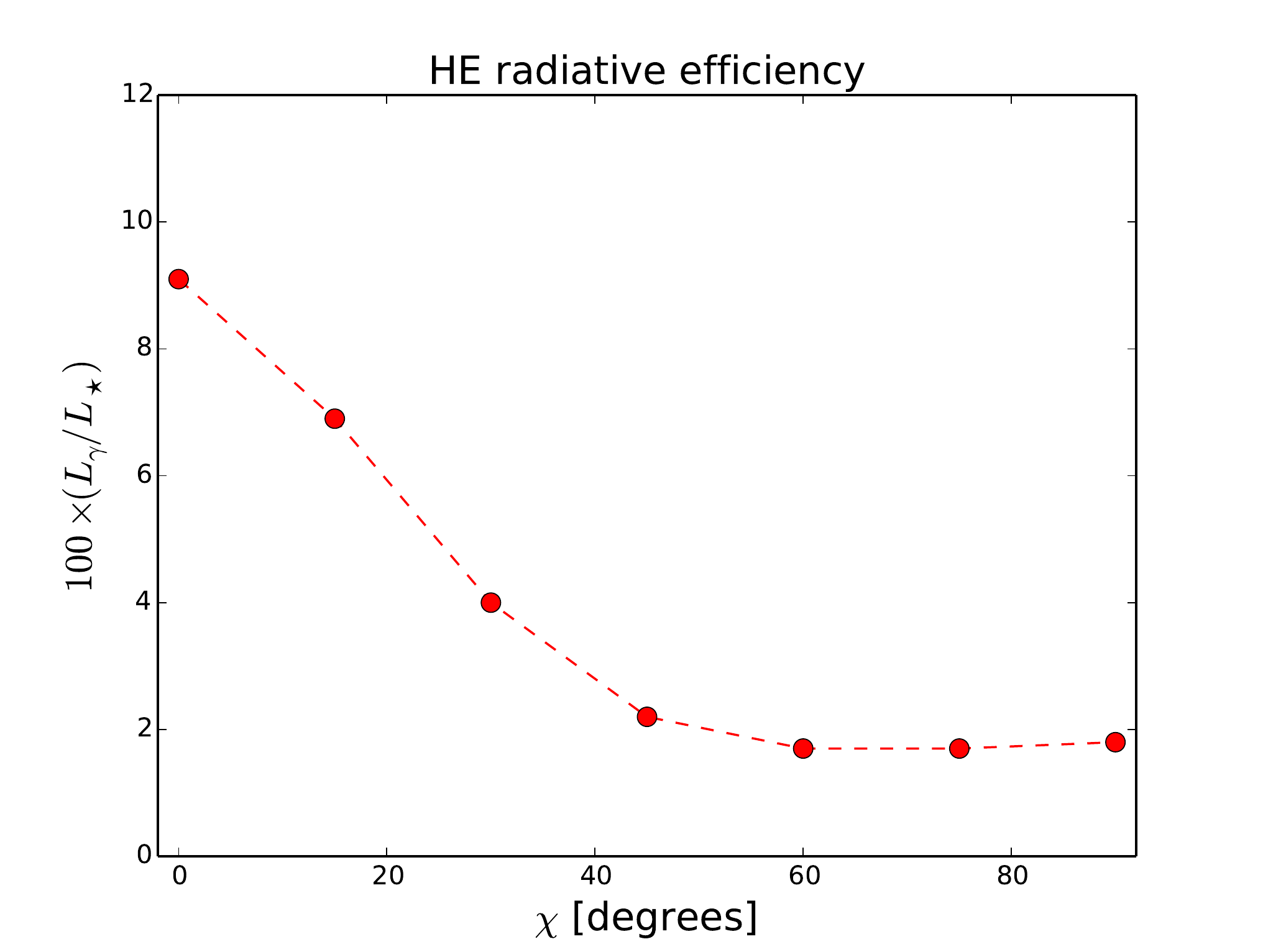}
\caption{{\bf Top: } Dissipated fraction of the total radial Poynting flux $L_{\star}$ (converted to non-thermal particles and radiation) within a sphere of radius $r=2R_{\rm LC}$ as a function of the pulsar inclination, as reported by global PIC studies \citep{2014ApJ...785L..33P, 2014ApJ...795L..22C, 2015MNRAS.448..606C, 2015ApJ...801L..19P, 2015MNRAS.449.2759B}. The values from \citet{2015MNRAS.448..606C} and \citet{2015MNRAS.449.2759B} are for their highest particle injection solutions (closest to force-free); lower particle injection leads to more dissipation ($20$-$50\%$). {\bf Bottom: } High-energy radiation efficiency reported by \citet{2016MNRAS.457.2401C} as a function of the pulsar inclination. The radiative efficiency is defined as the curvature and synchrotron radiative power summed over the numerical box and over all directions, $L_{\gamma}$, divided by the total radial Poynting flux, $L_{\star}$.}
\label{fig_dissipation}
\end{figure}
%%%%%%%%%%%%%%%%%%%%%%%%%%

The pulsar spindown reported by PIC simulations is in agreement with the ideal MHD expectations in the limit where abundant plasma is present in the magnetosphere (Figure~\ref{fig_spindown}). However, all PIC studies also show significant dissipation of the spindown power $L_\star$ inside and around the light cylinder. Figure~\ref{fig_dissipation} (top panel) shows the fraction of $L_\star$ that is dissipated within a sphere of radius $r=2R_{\rm LC}$ found in the literature, as a function of the pulsar inclination. The aligned rotator has the maximum dissipated fraction of about 20\%. The dissipation rate then decreases monotonically with increasing inclination to about 2.5\% for $\chi=90^{\rm o}$ \citep{2015ApJ...801L..19P}. The dissipated Poynting flux is not lost numerically, but instead it is self-consistently channelled into particle kinetic energy and non-thermal radiation. 

The scatter in $L/L_\star$ at $\chi=0^{\rm o}$ in Figure~\ref{fig_dissipation}
is most likely due to different prescriptions for particle injection. \citet{2015MNRAS.448..606C} did not implement $e^\pm$ creation in the magnetosphere and instead injected a mildly relativistic flow of $e^\pm$ from the stellar surface. This prescription tends to screen $E_\parallel$ and somewhat reduces dissipation. \citet{2015MNRAS.449.2759B} implemented $e^\pm$ creation proportionally to the local $E_\parallel$. \citet{2015ApJ...801L..19P} assumed instantaneous local pair creation by any particle reaching a fixed threshold Lorentz factor $\gamma_{\rm thr}$. \citet{2014ApJ...795L..22C} performed the most detailed simulation that followed the propagation of curvature gamma-rays and their conversion to pairs. All the simulations, however, show a qualitatively similar structure of the magnetosphere, in particular the formation of the Y-shaped current sheet with significant dissipation.

The dissipated fraction must increase as the pulsar slows down and approaches its ``deathline,'' when the pair creation rate declines so that it becomes increasingly difficult to screen $E_\parallel$. When \citet{2015MNRAS.448..606C} 
reduced plasma supply in their simulations, they observed that the magnetosphere becomes highly charge-separated with a low spindown power ($L_\star<L_0$,
some field lines remain closed outside the light-cylinder) and the dissipation fraction can reach $>50\%$. In this dissipative solution, the separatrix current layers and the Y-point disappear, and the equatorial current sheet is electrostatically supported (see also \citealt{2014ApJ...781...46C} solution).

\subsection{General relativistic effects}

In agreement with the criterion in \Eq~(\ref{eq:alpha}), \citet{2014ApJ...795L..22C} did not observe a polar-cap accelerator in the standard aligned-rotator problem --- the negative electric current flowing from the polar cap was carried by the mildly relativistic electron flow extracted from the star. \citet{2015ApJ...815L..19P} observed similar behavior of moderately inclined rotators, with inclinations $\chi<40^{\rm o}$. For larger inclinations, the polar cap became ``activated,'' in agreement with the analysis of the $\alpha$ parameter in force-free models: strongly oblique rotators have large variations of $\alpha$ across their polar caps (e.g., see Figure 1 in \citealt{2013MNRAS.429...20T}).

The lack of polar-cap activity at moderate inclinations may be problematic for explaining observed radio emission, in particular its ``core'' component that is usually associated with $e^\pm$ production inside the open field-line bundle. This motivated recent work on simulating general relativistic effects near pulsars, which change the parameter $\alpha$ by tens of percent \citep{2015ApJ...815L..19P, 2016arXiv160404625G, 2016arXiv160405670B}. The most important effect in this context is the frame-dragging of space-time by the stellar rotation \citep{1990SvAL...16..286B, 1992MNRAS.255...61M, 2003ApJ...584..427S}. The effective angular velocity of the magnetosphere is reduced by the Lense-Thirring angular velocity,
$\omega_{\rm LT}=(2/5)\Omega (r_{\rm g}/r_{\star})(r_{\star}/r)^3$,
where $r_{\rm g}=2GM_{\star}/c^2$ is the Schwarzschild radius. This leads to an increase of the $\alpha$ parameter,
\begin{equation}
\alpha_{\rm GR}\approx \frac{\alpha}{1-\omega_{\rm LT}/\Omega},
\end{equation}
where $\alpha$ on the right-hand side corresponds to the standard dipole rotator neglecting general relativistic effects. The effect is strongest at the stellar surface, where one finds $\alpha_{\rm GR}\approx 1.2\alpha$. \citet{2015ApJ...815L..19P} included the Lense-Thirring effect in the field solver (in a leading order) in a 2D axisymmetric PIC simulation, and observed the ignition of $e^\pm$ discharge above the polar cap for a sufficiently high compactness of the neutron star $M/\rs$. They also observed that the discharge operates in a quasi-periodic manner, in agreement with \citet{2008ApJ...683L..41B} and \citet{2013MNRAS.429...20T}, and may generate radio waves.

\subsection{Radiation from the equatorial current sheet}

The equatorial current sheet outside the light cylinder is a prominent dissipation site, especially if the magnetosphere itself is close to force-free (the pulsar is far from the deathline). The strong energy dissipation and particle acceleration in the equatorial sheet accompanies magnetic reconnection of the opposite open magnetic fluxes, which produces and ejects relativistic plasmoids \citep{2014ApJ...785L..33P, 2014ApJ...795L..22C, 2015MNRAS.448..606C}, as predicted by \citet{1990ApJ...349..538C} and \citet{1994ApJ...431..397M}. This dissipation occurs in the plasma outflow outside the light cylinder and may be studied in detail by PIC simulations even without an explicit implementation of pair creation. Instead, one can form a plasma-filled magnetosphere e.g. by sprinkling $e^\pm$ everywhere  \citep{2014ApJ...785L..33P} or by ejecting a flow of copious $e^\pm$ pairs from the stellar surface \citep{2015MNRAS.448..606C, 2016MNRAS.457.2401C}. This leads to the magnetic configuration close to the force-free solution with the Y-shaped current sheet, and to the strong dissipation in the equatorial plane outside the light cylinder.   
   
\begin{figure}[]
\centering
\includegraphics[width=5.8cm]{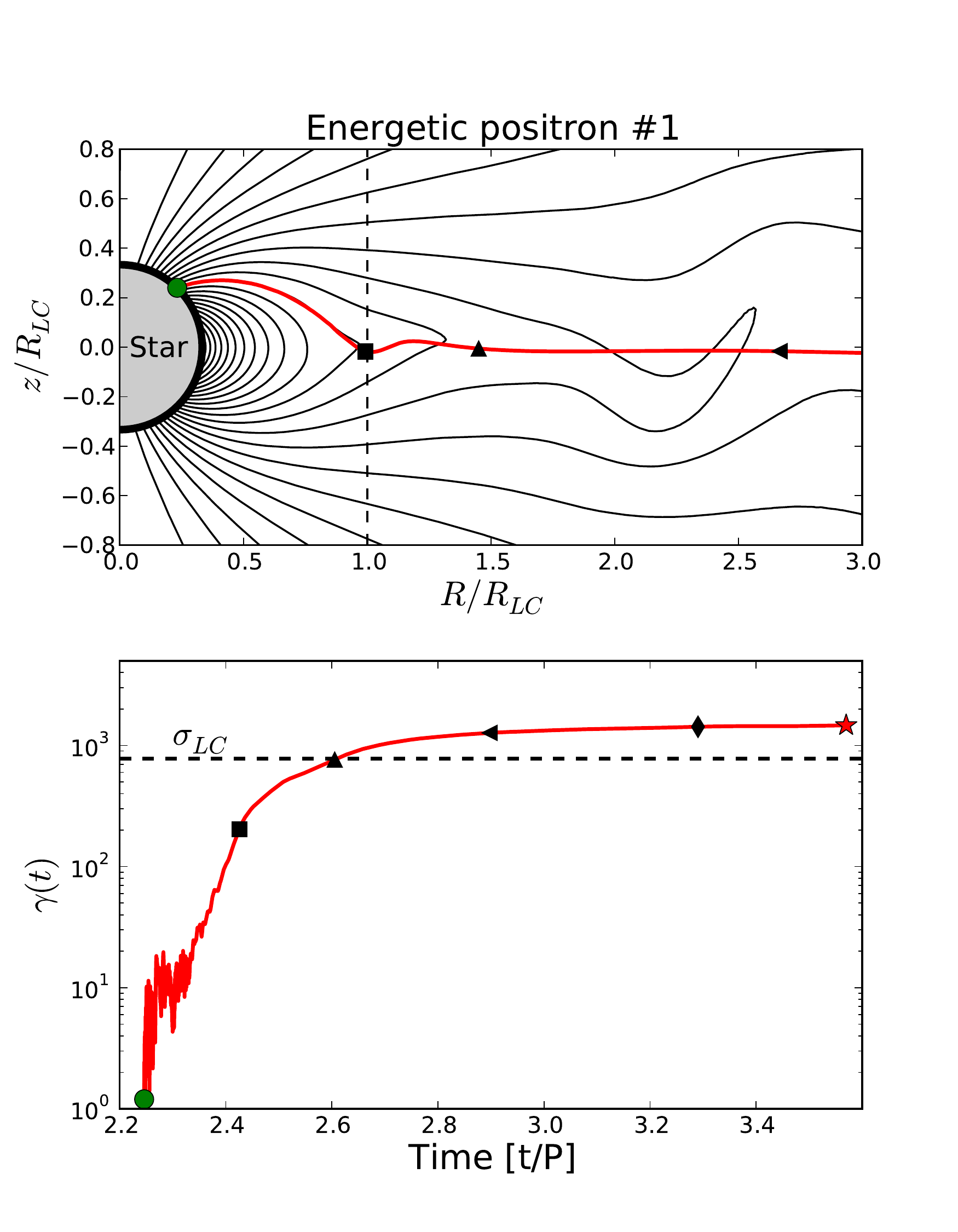}
\includegraphics[width=5.8cm]{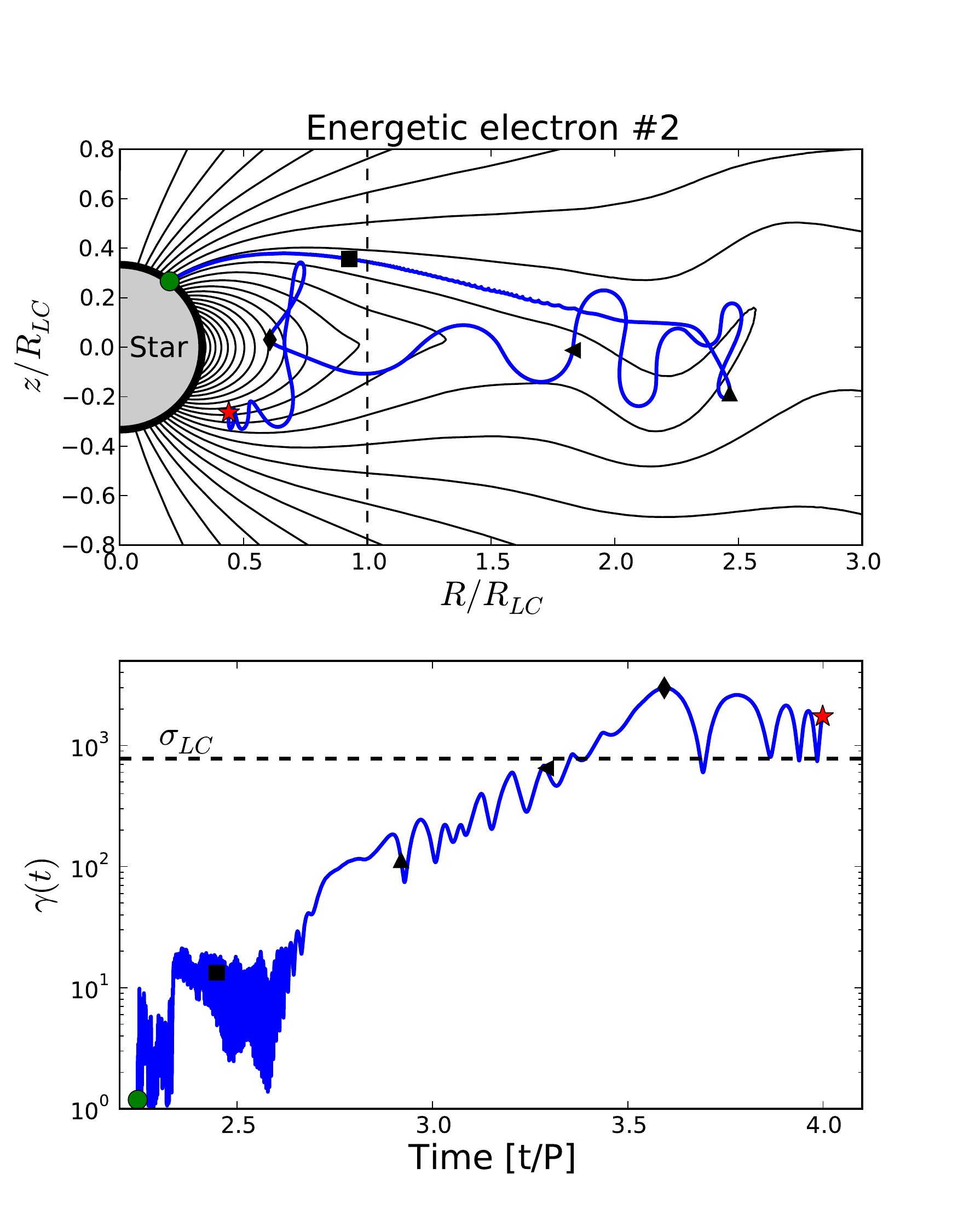}
\caption{{\bf Top:} Sample trajectories of a high-energy positron (left) and electron (right) accelerated in the current sheet of an aligned rotator, shown projected on the poloidal plane. {\bf Bottom:} The particle Lorentz factor as a
function of time. Black symbols on the curves indicate the correspondence between the time (bottom panels) and position on the trajectory (top panels). Figure adapted from \citet{2015MNRAS.448..606C}.}
\label{fig_orbit}
\end{figure}

An important parameter is the plasma magnetization defined as
\begin{equation}
\sigma=\frac{B^2}{4\pi \Gamma n m_{\rm e} c^2},
\end{equation}
where $n$ is the number density of the $e^\pm$ plasma and $\Gamma$ is its average bulk Lorentz factor. In pulsars, $\sigma\gg 1$ everywhere in the magnetosphere except the interior of the equatorial current sheet, and dissipation of a non-negligible fraction of the spindown power can significantly increase the energy per particle. Simulations show that most of the equatorial dissipation occurs near the light cylinder, around the Y-point up to a few $R_{\rm LC}$. The high-energy particles tend to pile up at an energy given by the magnetization of the plasma at the light cylinder, i.e. $\gamma\approx\sigma_{\rm LC}$ \citep{2014ApJ...785L..33P, 2015MNRAS.448..606C}. This energy should be compared with the maximum energy that a particle can get from the vacuum potential drop across the polar cap
% , i.e., 
$\phi_{\rm pc}=e\Phi_{\rm pc}/m_{\rm e}c^2$, where (using \Eqs~\ref{E_corotation}, \ref{eq_thetapc})
\begin{equation}
\Phi_{\rm pc}=\int_0^{\theta_{\rm pc}}E_{\theta}(R_{\star})R_{\star}d\theta=\frac{\mu\Omega^2}{c^2},
\end{equation}
so that \citep{2015MNRAS.448..606C}
\begin{equation}
\gamma\approx\sigma_{\rm LC}\sim\frac{\phi_{\rm pc}}{\kappa\Gamma},
\end{equation}
where $\kappa$ is the multiplicity of $e^\pm$ pairs ejected in the current sheet, defined relative to the minimum particle number required to sustain the electric current. Hence, the average particle ejected by a high-multiplicity pulsar earns only a small fraction of the vacuum potential drop. \citet{2015MNRAS.448..606C} tracked the motion of accelerated particles in their simulations of the aligned rotator and studied where the particles gain their energies. They found that positive charges are accelerated outward as they cross the Y-point region and get ejected along the equator. In contrast, electrons are mainly accelerated as they precipitate back towards the star along the current sheet (see Figure~\ref{fig_orbit}).

%%%%%%%%%%%%%%%%%%%%%%%%
%%%%%%%%%%%%%%%%%%%%%%%%

%%%%%%%%%%%%%%%%%%%%%%%%%%
\begin{figure}[]
\centering
\includegraphics[width=11cm]{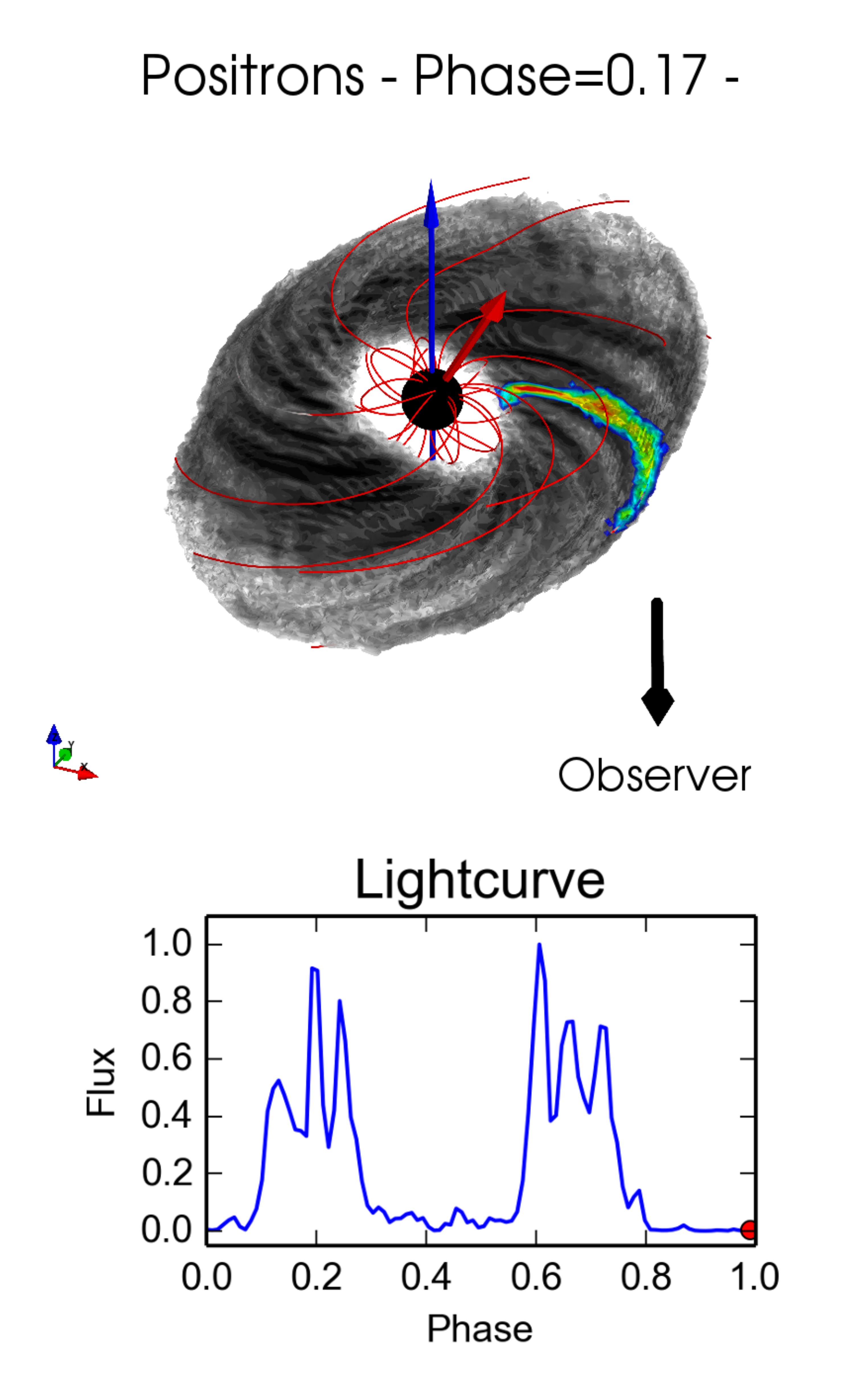}
\caption{{\bf Top:} Spatial distribution of the high-energy synchrotron radiation from an oblique rotator obtained with a 3D PIC simulation. The grey scale shows the isotropically integrated flux, while the color scale shows the emitting regions at the pulsar phase $0.17$ as seen by an observer looking along the equator. The angle between the rotation axis (blue arrow) and the magnetic axis (red arrow) is $\chi=30^{\rm o}$. Red curves are the magnetic field lines. {\bf Bottom:} Reconstructed high-energy pulse profile of radiation received by the observer. Figure adapted from \citet{2016MNRAS.457.2401C}.}
\label{fig_lc}
\end{figure}
%%%%%%%%%%%%%%%%%%%%%%%%%%

Efficient particle acceleration leads to strong non-thermal high-energy emission. The radiation-reaction force due to the curvature and synchrotron emission can be of the same order as the Lorentz force, and particle momentum perpendicular to the field lines may be quickly radiated away. These effects are self-consistently included in the PIC simulations. In classical electrodynamics, the radiation reaction force is given by the Landau-Lifshitz formula \citep{2010NJPh...12l3005T, 2016MNRAS.457.2401C}
\begin{eqnarray}
\mathbf{g} =\frac{2}{3}r^2_{\rm e}\left[\left(\mathbf{E}+\boldsymbol{\beta}\times\mathbf{B}\right)\times\mathbf{B}+\left(\boldsymbol{\beta}\cdot\mathbf{E}\right)\mathbf{E}\right]\hspace{2cm}\nonumber \\
-\frac{2}{3}r^2_{\rm e}\gamma^2\left[\left(\mathbf{E}+\boldsymbol{\beta}\times\mathbf{B}\right)^2-\left(\boldsymbol{\beta}\cdot\mathbf{E}\right)^2\right]\boldsymbol{\beta},
\label{frad}
\end{eqnarray}
where $r_{\rm e}$ is the classical radius of the electron. The 3D radiative PIC simulations of \citet{2016MNRAS.457.2401C} showed that the strong curvature and synchrotron cooling had little impact on the overall structure of the oblique rotator and its spindown rate (Figure~\ref{fig_spindown}). Radiative cooling also has little impact on the maximum energy reached by the particles accelerated in the current sheet, because the particles are accelerated while
being focused deep inside the layer where the perpendicular magnetic field virtually vanishes \citep{2004PhRvL..92r1101K, 2011ApJ...737L..40U, 2013ApJ...770..147C}. However, when the particles leave or re-enter the layer
they radiate efficiently. The total isotropic power radiated away in the form of synchrotron radiation reaches almost 10\% of the spindown for the aligned pulsar and decreases down to a few percent at high inclinations (bottom panel in Figure~\ref{fig_dissipation}), because the dissipation rate decreases with inclination. Such radiative efficiencies are consistent with observed high-energy gamma-ray efficiencies reported by the {\em Fermi}-LAT \citep{2010ApJS..187..460A, 2013ApJS..208...17A}.

The expected radiative signatures --- pulse profiles and spectra --- can then be calculated from the instantaneous synchrotron and curvature radiation spectrum emitted by each simulation particle,
\begin{eqnarray}
F_{\nu}\left(\nu\right) &=& \frac{\sqrt{3}e^3 \tilde{B}_{\perp}}{m_{\rm e}c^2}\left(\frac{\nu}{\nu_{\rm c}}\right)\int_{\nu/\nu_{\rm c}}^{+\infty}K_{5/3}(x)dx \\
\nu_{\rm c} &=& \frac{3 e \tilde{B}_{\perp}\gamma^2}{4\pi m_{\rm e}c}\\
\tilde{B}_{\perp} &=& \sqrt{\left(\mathbf{E}+\boldsymbol{\beta}\times\mathbf{B}\right)^2-\left(\boldsymbol{\beta}\cdot\mathbf{E}\right)^2}.
\end{eqnarray}
This is the usual formula (e.g. \citealt{1970RvMP...42..237B}) except that the effective perpendicular magnetic field, $\tilde{B}_{\perp}$ here includes the electric field, because it is of the same order as $B$ in pulsar magnetospheres. Photons are then collected on a spherical screen located at infinity and they are folded over the pulsar phase to reconstruct the pulse profiles.

The simulations show that the strongest high-energy emission is produced in the equatorial current in the form of synchrotron photons \citep{2016MNRAS.457.2401C}, as envisioned by \citet{1996A&A...311..172L}. Additional high-energy radiation is produced in the separatrix current sheets inside the light cylinder \citep{2014ApJ...795L..22C}. The simulated high-energy pulse profiles typically have two peaks, in agreement with gamma-ray observations \citep{2010ApJS..187..460A, 2013ApJS..208...17A}. Each peak appears when the equatorial current sheet crosses the observer's line of sight (Figure~\ref{fig_lc}, \citealt{2016MNRAS.457.2401C}). The form of the pulse profile (i.e., phase-separation, amplitude and width of pulses) is entirely determined by the geometrical shape of the current sheet.

These studies prelude the beginning of direct comparison between PIC simulations and observations. As a proof of principle, \citet{2016MNRAS.463L..89C} showed that polarization modelling along with pulse profile fitting can provide independent constraints on the pulsar inclination and viewing angles. Applied to the Crab pulsar, they find an obliquity angle $\chi=60^{\rm o}$ and viewing angle $\alpha=130^{\rm o}$ which are in good agreement with values inferred from the morphology of the pulsar wind nebula in X-rays \citep{2004ApJ...601..479N, 2012ApJ...746...41W}. A more detailed and systematic analysis extended to other pulsars should be performed in the future.

\section{Summary}\label{sect_ccl}

The theory of pulsar electrodynamics is complex. Although the physics problem is well posed, it entangles several processes such as particle acceleration, non-thermal emission, pair creation, and relativistic reconnection, which make the problem hard. The force-free electrodynamics  and MHD models demonstrate some of the essential elements of the pulsar magnetosphere --- the spindown power, the magnetic topology, and the pattern of electric currents --- and highlight the presence of a prominent Y-shaped current sheet. The recent PIC numerical experiments, which attempt to solve the full problem from first principles, have been a game changer in pulsar theory. They show how the magnetosphere is filled with plasma and how a fraction of the spindown power is dissipated and deposited into high-energy particles that produce observed radiation. For the first time, numerical simulations became capable of predicting observables from first principles, opening new perspectives for our understanding of pulsars. 

\begin{acknowledgements}
The authors thank the International Space Science Institute in Bern for the invitation and hospitality during the Workshop on Jets and Winds in
Pulsar Wind Nebulae and Gamma-ray Bursts in  November 2015. BC acknowledges support from CNES and Labex OSUG@2020 (ANR10 LABX56). AMB pulsar research is supported by NASA grants NNX13AI34G and NNX15AU71G.
\end{acknowledgements}

\bibliographystyle{aps-nameyear}          % American Physical Society (APS) style, author-year citations
\bibliography{ISSI_pulsar}                % name your BibTeX data base
\nocite{*}

\end{document}